\documentclass[sigconf]{acmart}
\usepackage{multirow}
\usepackage{float}
\usepackage{subfigure}
\usepackage{enumerate}
\usepackage{enumitem}
\usepackage{stfloats}
\usepackage{hyperref}
\usepackage{breakurl}
\usepackage{diagbox}
\usepackage{balance}
\usepackage{titlesec}
\usepackage{verbatimbox}

\usepackage{ulem}
\usepackage{indentfirst}
\usepackage[linesnumbered,ruled,vlined]{algorithm2e}
\usepackage{cleveref}
\usepackage{colortbl}

\crefname{section}{§}{§§}

\definecolor{best}{RGB}{255, 159, 69}
\definecolor{gray}{RGB}{221, 221, 221}

\AtBeginDocument{%
  \providecommand\BibTeX{{%
    \normalfont B\kern-0.5em{\scshape i\kern-0.25em b}\kern-0.8em\TeX}}}

\copyrightyear{2022}
\acmYear{2022}
\setcopyright{acmcopyright}
\acmConference[KDD '22] {Proceedings of the 28th ACM SIGKDD Conference on Knowledge Discovery and Data Mining}{August 14--18, 2022}{Washington, DC, USA.}
\acmBooktitle{Proceedings of the 28th ACM SIGKDD Conference on Knowledge Discovery and Data Mining (KDD '22), August 14--18, 2022, Washington, DC, USA}
\acmPrice{15.00}
\acmISBN{978-1-4503-9385-0/22/08}
\acmDOI{10.1145/3534678.3539452}

\def\cyk{\color{black} }

\def\emb{\boldsymbol} 
\DeclareMathOperator\sign{sign}
\DeclareMathOperator\argmin{argmin}

\begin{document}
\fancyhead{}
\title{Learning Binarized Graph Representations with Multi-faceted Quantization Reinforcement for Top-K Recommendation}


\author{
Yankai Chen$^1$,
Huifeng Guo$^2$,
Yingxue Zhang$^2$,
Chen Ma$^3$,
Ruiming Tang$^2$,
Jingjie Li$^2$,
Irwin King$^1$
}
 \affiliation{
  \institution{$^1$The Chinese University of Hong Kong, $^2$Huawei Noah's Ark Lab, $^3$City University of Hong Kong}
  \city{}
  \country{}
}
  \email{{ykchen, king}@cse.cuhk.edu.hk; chenma@cityu.edu.hk; {huifeng.guo, yingxue.zhang, tangruiming, lijingjie1}@huawei.com }

\begin{abstract}

\noindent Learning vectorized embeddings is at the core of various recommender systems for user-item matching.
{\cyk To perform efficient online inference,} \textit{representation quantization}, aiming to embed the latent features by a compact sequence of discrete numbers, recently shows the promising potentiality in optimizing both memory and computation overheads.
However, existing work merely focuses on \textit{numerical quantization} whilst ignoring the concomitant \textit{information loss} issue, which, consequently, leads to conspicuous performance degradation.
In this paper, we propose a novel quantization framework to learn \textit{\underline{Bi}narized} \textit{\underline{G}raph} \textit{R\underline{e}present\underline{a}tions} for \textit{Top-K} \textit{\underline{R}ecommendation} (BiGeaR).
BiGeaR introduces multi-faceted quantization reinforcement at the \textit{pre-}, \textit{mid-}, and \textit{post-stage} of {\cyk binarized} representation learning, which substantially retains the representation informativeness against embedding binarization.
In addition to saving the memory footprint, BiGeaR further develops solid online inference acceleration with bitwise operations, providing alternative flexibility for the realistic deployment.
The empirical results over five large real-world benchmarks show that BiGeaR achieves about 22\%$\sim$40\% performance improvement over the state-of-the-art  quantization-based recommender system, and recovers about 95\%$\sim$102\% of the performance capability of the best full-precision counterpart with over 8$\times$ time and space reduction.
\end{abstract}



\maketitle

\section{Introduction}
Recommender systems, aiming to perform personalized information filtering~\cite{pinsage}, are versatile to many Internet applications. 
Learning vectorized user-item representations (i.e., embeddings), as the core of various recommender models, is the prerequisite for online inference of user-item interactions~\cite{lightgcn,ngcf}.
With the explosive data expansion (e.g., Amazon owns over 150M users and 350M products~\cite{amazon,amazon_2}), one major existing challenge however is to perform inference efficiently.
This usually requires large memory and computation consumption (e.g., for Amazon 500M-scaled \textit{full-precision}\footnote{{\cyk It could be single-precision or double-precision;  we use float32 as the default for explanation and conducting experiments throughout this paper.}} embedding table) on certain data centers~\cite{tailor2020degree}, and therefore hinders the deployment to devices with limited resources~\cite{tailor2020degree}.

 \begin{figure}[tp]
\begin{minipage}{0.5\textwidth}
\hspace{-0.2in}
\includegraphics[width=3.5in]{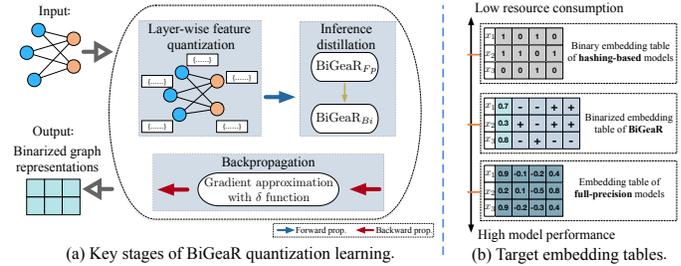}
\end{minipage} 
\setlength{\abovecaptionskip}{0.2cm}
\setlength{\belowcaptionskip}{0.2cm}
\vspace{-0.05in}
\caption{Illustration of BiGeaR.}
\label{fig:intro}
\end{figure}

To tackle this issue, \textit{representation quantization} recently provides the promising feasibility.
Generally, it learns to quantize latent features of users and items via converting the continuous full-precision representations into discrete low-bit ones.
The quantized representations thus are conducive to model size reduction and inference speed-up with low-bit arithmetic on devices where CPUs are typically more affordable than expensive GPUs~\cite{banner2018scalable,bahri2021binary}.
Technically, quantization can be categorized into multi-bit, 2-bit (i.e., ternarized), and 1-bit (i.e., binarized) quantization. 
With only one bit-width, representation binarization for recommendation takes the most advantage of representation quantization and therefore draws the growing attention recently~\cite{hashgnn,kang2019candidate}.

Despite the promising potentiality, it is still challenging to develop realistic deployment mainly because of the large performance degradation in Top-K recommendation~\cite{hashgnn,kang2019candidate}.
The crux of the matter is the threefold \textit{information loss}:
\begin{itemize}[leftmargin=*,topsep=2pt,parsep=0.5pt]
\item \textbf{Limited expressivity of latent features.}
Because of the discrete constraints, mapping full-precision embeddings into compact binary codes with equal expressivity is NP-hard~\cite{haastad2001some}.
{\cyk Thus, instead of proposing complex and deep neural structures for quantization~\cite{erin2015deep,zhu2016deep}, $\sign(\cdot)$ function is widely adopted to achieve {\small $O(1)$} embedding binarization~\cite{hashgnn,wang2017survey,lin2017towards}.} 
However, this only guarantees the sign (+/-) correlation for each embedding entry.
{\cyk Compared to the original full-precision embeddings, binarized targets produced from $\sign(\cdot)$ are naturally less informative}.

\item \textbf{Degraded ranking capability.}
Ranking capability, as the essential measurement of Top-K recommendation, is the main objective to work on. 
{\cyk Apart from the inevitable feature loss in numerical quantization, previous work further ignores the discrepancy of hidden knowledge that is inferred by full-precision and binarized embeddings~\cite{kang2019candidate,hashgnn}.
However, such hidden knowledge is vital to reveal users' preference towards different items, losing of which may thus draw degraded ranking capability and sub-optimal model learning accordingly.
}

\item \textbf{Inaccurate gradient estimation.}
Due to the non-differentiability of quantization function $\sign(\cdot)$, \textit{Straight-Through} \textit{Estimator} (STE) \cite{bengio2013estimating} is widely adopted to assume all propagated gradients as 1 in backpropagation~\cite{hashgnn,lin2017towards,qin2020forward}.
Intuitively, the integral of 1 is a certain linear function other than $\sign(\cdot)$, whereas this may lead to inaccurate gradient estimation and produce inconsistent optimization directions in the model training.
\end{itemize}


To address aforementioned problems, we propose a novel quantization framework, namely \textbf{BiGeaR}, to learn the \textit{\underline{Bi}narized} \textit{\underline{G}raph} \textit{R\underline{e}present\underline{a}tions} for \textit{Top-K} \textit{\underline{R}ecommendation}.
Based on the natural bipartite graph topology of user-item interactions, we implement BiGeaR with the inspiration from graph-based models, i.e., Graph Convolutional Networks (GCNs)~\cite{kipf2016semi,graphsage}.
With the emphasis on deepening the exploration of multi-hop subgraph structures,  GCN-based recommender models capture the high-order interactive relations and well simulate \textit{Collaborative} \textit{Filtering} for recommendation~\cite{lightgcn,ngcf,pinsage}.
Specifically, BiGeaR sketches graph nodes (i.e., users and items) with binarized representations, which facilitates nearly one bit-width representation compression.
Furthermore, our model BiGeaR decomposes the prediction formula (i.e., \textit{inner} \textit{product}) into bitwise operations (i.e., {\tt Popcount} and {\tt XNOR}).
This dramatically reduces the number of floating-point operations (\#FLOP) and thus introduces theoretically solid acceleration {\cyk for online inference}.
{\cyk To avoid large information loss}, as shown in Figure~\ref{fig:intro}(a), BiGeaR technically consists of multi-faceted quantization reinforcement at the \textit{pre}-, \textit{mid}-, and \textit{post}-stage of binarized representation learning:
\begin{enumerate}[leftmargin=*,topsep=2pt,parsep=0.5pt]
\item At the pre-stage of model learning, we propose the \textbf{graph} \textbf{layer-wise} \textbf{quantization} {\cyk (from low- to high-order interactions)} to consecutively binarize the user-item features with different semantics. 
Our analysis indicates that such layer-wise quantization can actually achieve the \textit{magnification} \textit{effect} of \textit{feature} \textit{uniqueness},
{\cyk which significantly compensates for the limited expressivity of embeddings binarization.}
The empirical study also justifies that, this is more effective to enrich the quantization informativeness, rather than simply increasing the embedding sizes in the conventional manner~\cite{hashgnn,lin2017towards,qin2020forward,rastegari2016xnor}.

\item During the mid-stage of embedding quantization, BiGeaR introduces the \textbf{self-supervised inference distillation} to develop the \textit{low-loss} ranking capability inheritance.
Technically, it firstly extracts several \textit{pseudo-positive} \textit{training} \textit{samples} that are inferred by full-precision embeddings of BiGeaR.
Then these samples {\cyk serve} as the direct regularization target to the quantized embeddings, such that they {\cyk can distill the ranking knowledge from full-precision ones to have similar inference results}.
Different from the conventional knowledge distillation, our approach is tailored specifically for Top-K recommendation and therefore boosts the performance {\cyk with acceptable training costs}.

\item As for the post-stage backpropagation of quantization optimization, we propose to utilize the approximation of \textit{Dirac} \textit{delta} \textit{function} (i.e., $\delta$ function)~\cite{bracewell1986fourier} for more \textbf{accurate gradient estimation}. 
In contrast to STE, our gradient estimator provides the consistent optimization direction with $\sign(\cdot)$ in both forward and backward propagation. 
The empirical study demonstrates its superiority over other gradient estimators.
\end{enumerate}

\noindent\textbf{Empirical Results.}
The extensive experiments over five real-world benchmarks show that, BiGeaR significantly outperforms the state-of-the-art quantization-based recommender model by 25.77\%$\sim$40.12\% and 22.08\%$\sim$39.52\% \textit{w.r.t.} Recall and NDCG metrics.
Furthermore, it attains 95.29\%$\sim$100.32\% and 95.32\%$\sim$101.94\% recommendation capability compared to the best-performing full-precision model, {\cyk with over 8$\times$ inference acceleration and space compression}. 


\noindent\textbf{Discussion.} 
It is worthwhile mentioning that BiGeaR is related to \textit{hashing-based} models (i.e., learning to hash)~\cite{kang2021learning,kang2019candidate}, {\cyk as, conceptually, binary hashing can be viewed as 1-bit quantization}. However, as shown in Figure~\ref{fig:intro}(b), they have different {\cyk motivations}.
Hashing-based models are usually designed for fast candidate generation, followed by full-precision \textit{re-ranking} algorithms for accurate prediction.
Meanwhile, BiGeaR is \textit{end-to-end} that aims to make predictions within the proposed architecture.
Hence, we believe BiGeaR is \textit{technically} \textit{related} but \textit{motivationally} \textit{orthogonal} to them.

\noindent\textbf{Organization.}
We present BiGeaR methodology and model analysis in~\cref{sec:end} and~\cref{sec:analysis}. 
Then we report the experiments and review the related work in~\cref{sec:exp} and~\cref{sec:work} with the conclusion in ~\cref{sec:con}.

\section{BiGeaR Methodology}
\label{sec:end}
In this section, we formally introduce:
\textit{(1) graph layer-wise quantization for feature magnification;
(2) inference distillation for ranking capability inheritance;
(3) gradient estimation for better model optimization.}
BiGeaR framework is illustrated in Figure~\ref{fig:framework}(a).
The notation table and pseudo-codes are attached in Appendix~\ref{app:notation} and~\ref{app:algo}.

\begin{figure*}[tp]
\begin{minipage}{1\textwidth}
\includegraphics[width=7in]{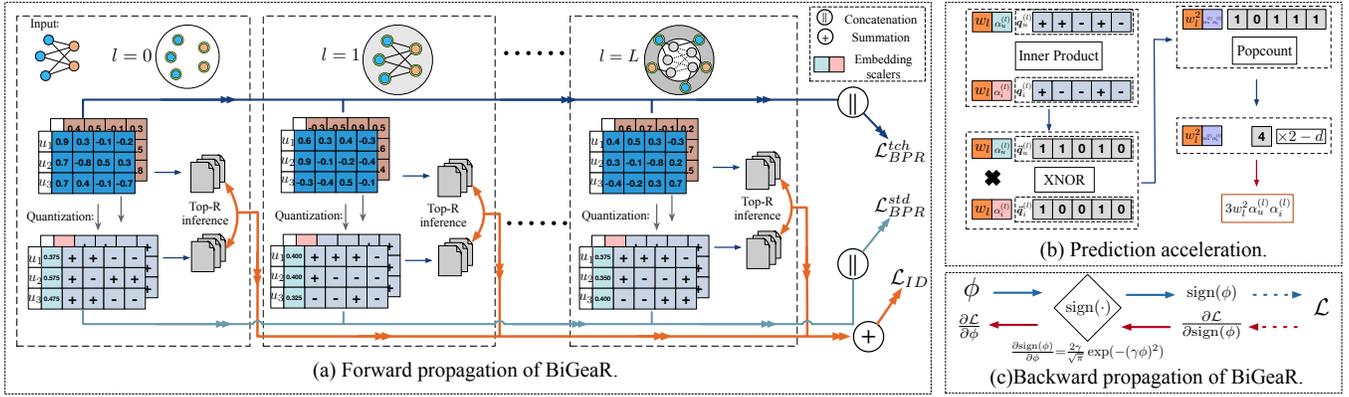}
\end{minipage} 
\setlength{\abovecaptionskip}{0.2cm}
\setlength{\belowcaptionskip}{0.2cm}
\vspace{-0.05in}
\caption{BiGeaR first pre-trains the full-precision embeddings and then triggers the (1) graph layer-wise quantization, (2) inference distillation, and (3) accurate gradient estimation to learn the binarized representations (Best view in color).}
\label{fig:framework}
\end{figure*}


\textbf{Preliminaries: graph convolution.} Its general idea is to learn node representations by iteratively propagating and aggregating latent features of neighbors via the graph topology~\cite{wu2019simplifying,lightgcn,kipf2016semi}. 
We adopt the graph convolution paradigm working on the continuous space from LightGCN~\cite{lightgcn} that recently shows good recommendation performance.
Let {\small$\boldsymbol{v}^{(l)}_u, \boldsymbol{v}^{(l)}_i\in \mathbb{R}^{d}$} denote the continuous feature embeddings of user $u$ and item $i$ computed after $l$ layers of information propagation. 
{\small $\mathcal{N}(x)$} represents $x$'s neighbor set.
They can be iteratively updated by utilizing information from the ($l$-$1$)-th layer: 
\begin{small}
\begin{equation}
\label{eq:gcn}
\setlength\abovedisplayskip{2pt}
\setlength\belowdisplayskip{2pt}
\resizebox{0.9\linewidth}{!}{$
\displaystyle
\boldsymbol{v}^{(l)}_u = \sum_{i\in \mathcal{N}(u)} \frac{1}{\sqrt{|\mathcal{N}(u)|\cdot|\mathcal{N}(i)|}}\boldsymbol{v}^{(l-1)}_i, \ \ \boldsymbol{v}^{(l)}_i = \sum_{u\in \mathcal{N}(i)} \frac{1}{\sqrt{|\mathcal{N}(i)|\cdot|\mathcal{N}(u)|}}\boldsymbol{v}^{(l-1)}_u.
$}
\end{equation}
\end{small}%

\vspace{-0.1in}
\subsection{\textbf{Graph Layer-wise Quantization}}
We propose the \textit{graph layer-wise quantization} mainly by computing \textbf{quantized embeddings} and \textbf{embedding scalers}:
(1) these quantized embeddings sketch the full-precision embeddings with $d$-dimensional binarized codes (i.e., $\{-1, 1\}^d$);
and (2) each embedding scaler reveals the value range of original embedding entries. 
Specifically, during the graph convolution at each layer, we track the intermediate information (e.g., {\small $\boldsymbol{v}^{(l)}_u$}) and perform the layer-wise 1-bit quantization in parallel as:
\begin{small}
\begin{equation}
\setlength\abovedisplayskip{2pt}
\setlength\belowdisplayskip{2pt}
\boldsymbol{q}_u^{(l)} = \sign\big(\boldsymbol{v}^{(l)}_u\big), \ \ \boldsymbol{q}_i^{(l)} = \sign\big(\boldsymbol{v}^{(l)}_i\big),
\end{equation}
\end{small}%
where embedding segments {\small$\boldsymbol{q}_u^{(l)}$, $\boldsymbol{q}_i^{(l)}$ $\in$ $\{-1, 1\}^d$} retain the node latent features directly from {\small$\boldsymbol{v}^{(l)}_u$} and {\small$\boldsymbol{v}^{(l)}_i$}.
To equip with the layer-wise quantized embeddings, we further include a layer-wise positive embedding scaler for each node (e.g., {\small$\alpha_u^{(l)}$ $\in$ $\mathbb{R}^+$}), such that {\small$\boldsymbol{v}^{(l)}_u$ $\doteq$ $\alpha_u^{(l)}$$\boldsymbol{q}^{(l)}_u$}. Then for each entry in {\small$\alpha_u^{(l)}$$\boldsymbol{q}^{(l)}_u$}, it is still binarized by {\small$\{-\alpha_u^{(l)}$}, {\small$\alpha_u^{(l)}\}$}.
In this work, we compute the mean of L1-norm as:
\begin{small}
\begin{equation}
\setlength\abovedisplayskip{2pt}
\setlength\belowdisplayskip{2pt}
\alpha_u^{(l)} = \frac{1}{d}\cdot||\boldsymbol{v}_u^{(l)}||_1, \ \ \alpha_i^{(l)} = \frac{1}{d} \cdot ||\boldsymbol{v}_i^{(l)}||_1.
\end{equation}
\end{small}%
Instead of setting {\small $\alpha_u^{(l)}$} and {\small $\alpha_i^{(l)}$} as learnable, such \textit{deterministic} computation is simple yet effective to provide the scaling functionality whilst substantially pruning the parameter search space. The experimental demonstration is in Appendix~\ref{app:scaler}. 

After $L$ layers of quantization and scaling, we have built the following \textbf{binarized embedding table} for each graph node $x$ as:
\begin{small}
\begin{equation}
\setlength\abovedisplayskip{2pt}
\setlength\belowdisplayskip{2pt}
\mathcal{A}_x = \{\alpha_x^{(0)}, \alpha_x^{(1)}, \cdots, \alpha_x^{(L)}\}, \ \ \mathcal{Q}_x = \{\boldsymbol{q}^{(0)}_x, \boldsymbol{q}^{(1)}_x, \cdots, \boldsymbol{q}^{(L)}_x\}.
\end{equation}
\end{small}%
From the technical perspective, BiGeaR binarizes the intermediate semantics at different layers of {\cyk the receptive field~\cite{velivckovic2017graph,wu2020comprehensive} for each node}.
This, essentially, achieves the \textbf{magnification effect of feature uniqueness} to enrich user-item representations via the interaction graph exploration. We leave the analysis in~\cref{sec:necessity}.

\vspace{-0.1in}
\subsection{Prediction Acceleration}
\textbf{Model Prediction.}
Based on the learned embedding table, we predict the matching scores by adopting the inner product:
\begin{small}
\begin{equation}
\setlength\abovedisplayskip{2pt}
\setlength\belowdisplayskip{2pt}
\label{eq:score}
\widehat{y}_{u,i} =  \big<f(\mathcal{A}_u, \mathcal{Q}_u), f(\mathcal{A}_i, \mathcal{Q}_i)\big>,
\end{equation}
\end{small}%
where function $f(\cdot, \cdot)$ in this work is implemented as:  
\begin{small}
\begin{equation}
\setlength\abovedisplayskip{2pt}
\setlength\belowdisplayskip{2pt}
\label{eq:useQ}
f(\mathcal{A}_u, \mathcal{Q}_u) = \Big|\Big|_{l=0}^L w_l  \alpha_u^{(l)} \boldsymbol{q}^{(l)}_u, \ \ f(\mathcal{A}_i, \mathcal{Q}_i) = \Big|\Big|_{l=0}^L w_l \alpha_i^{(l)} \boldsymbol{q}^{(l)}_i. 
\end{equation}
\end{small}%
Here $\big|\big|$ represents concatenation of binarized embedding segments, in which weight $w_l$ measures the contribution of each segment in prediction.
$w_l$ can be defined as a hyper-parameter or a learnable variable (e.g., with attention mechanism~\cite{velivckovic2017graph}).
In this work, we set $w_l$ $\propto$ $l$ to linearly increase $w_l$ for segments from lower-layers to higher-layers, mainly for its computational simplicity and stability.

\textbf{Computation Acceleration.}
Notice that for each segment of {\small $f(\mathcal{A}_u, \mathcal{Q}_u)$}, e.g., {\small $w_l\alpha_u^{(l)} \boldsymbol{q}^{(l)}_u$}, entries are binarized by two values (i.e., {\small$-w_l  \alpha_u^{(l)}$} or {\small$w_l  \alpha_u^{(l)}$}). 
Thus, we can achieve the prediction acceleration by decomposing Equation~(\ref{eq:score}) with \textit{bitwise operations}. 
Concretely, in practice, {\small$\boldsymbol{q}^{(l)}_u$} and {\small$\boldsymbol{q}^{(l)}_i$} will be firstly encoded into basic $d$-bits binary codes, denoted by {\small$\boldsymbol{{\ddot q}}^{(l)}_u, \boldsymbol{{\ddot q}}^{(l)}_i \in \{0,1\}^d$}.
Then we replace Equation~(\ref{eq:score}) by introducing the following formula:
\begin{small}
\begin{equation}
\label{eq:bit}
\setlength\abovedisplayskip{2pt}
\setlength\belowdisplayskip{2pt}
\widehat{y}_{u,i} = \sum_{l=0}^{L} w_l^2\alpha_u^{(l)}\alpha_i^{(l)}\cdot \big(2{\tt Popcount} \big({\tt XNOR}(\boldsymbol{{\ddot q}}^{(l)}_u, \boldsymbol{{\ddot q}}^{(l)}_i  )\big) - d\big).
\end{equation}
\end{small}%
Compared to the original computation approach in Equation~(\ref{eq:score}), our bitwise-operation-supported prediction in Equation~(\ref{eq:bit}) reduces the number of floating-point operations (\#FLOP) with {\tt Popcount} and {\tt XNOR}. We illustrate an example in Figure~\ref{fig:framework}(b).

\vspace{-0.1in}
\subsection{Self-supervised Inference Distillation}
To alleviate the \textit{asymmetric inference capability} issue between full-precision representations and binarized ones, we introduce the \textit{self-supervised inference distillation} such that binarized embeddings can well inherit the inference knowledge from full-precision ones.
Generally, we treat our full-precision intermediate embeddings (e.g., {\small $\boldsymbol{v}_u^{(l)}$}) as the \textbf{teacher} embeddings and the quantized segments as the \textbf{student} embeddings.
Given both teacher and student embeddings, we can obtain their prediction scores as {\small $\widehat{{y}}_{u,i}^{tch}$} and {\small $\widehat{{y}}_{u,i}^{std}$}.
For Top-K recommendation, then our target is to reduce their discrepancy as:
\begin{small}
\begin{equation}
\setlength\abovedisplayskip{2pt}
\setlength\belowdisplayskip{2pt}
\argmin \mathcal{D}(\widehat{{y}}_{u,i}^{tch}, \widehat{{y}}_{u,i}^{std}).
\end{equation}
\end{small}%
A straightforward implementation of function $\mathcal{D}$ from the conventional \textit{knowledge distillation}~\cite{hinton2015distilling,anil2018large} is to minimize their Kullback-Leibler divergence (KLD) or mean squared error (MSE). 
Despite their effectiveness in classification tasks (e.g., visual recognition~\cite{anil2018large,xie2020self}), they may not be appropriate for Top-K recommendation as:
\begin{itemize}[leftmargin=*,topsep=2pt,parsep=0.5pt]
\item Firstly, both KLD and MSE in $\mathcal{D}$ encourage the student logits (e.g., {\small $\widehat{{y}}_{u,i}^{std}$}) to be similarly distributed with the teacher logits in a macro view. But for ranking tasks, they may not well learn the relative order of user preferences towards items, which, however, is the key to improving Top-K recommendation capability.

\item Secondly, they both develop the distillation over the whole item corpus, which may be computational excessive for realistic model training.
As the item popularity usually follows the Long-tail distribution~\cite{park2008long,tang2018ranking}, learning the relative order of those frequently interacted items located at the tops of ranking lists actually contributes more to the Top-K recommendation performance.
\end{itemize}

To develop effective inference distillation, we propose to extract additional \textit{pseudo-positive training samples} from teacher embeddings to regularize the targeted embeddings on each convolutional layer. 
Concretely, let $\sigma$ represent the activation function (e.g., Sigmoid). We first pre-train the \textbf{teacher} embeddings to minimize the \textit{Bayesian Personalized Ranking} (BPR) loss~\cite{rendle2012bpr}:
\begin{small}
\begin{equation}
\setlength\abovedisplayskip{2pt}
\setlength\belowdisplayskip{-2pt}
\label{eq:hd-bpr}
\mathcal{L}^{tch}_{BPR} = -\sum_{u \in \mathcal{U}} \sum_{i\in \mathcal{N}(u) \atop j\notin \mathcal{N}(u)} \ln \sigma(\widehat{y}^{\,tch}_{u,i} - \widehat{y}^{\,tch}_{u,j}),
\end{equation}
\end{small}%
where {\small $\mathcal{L}^{tch}_{BPR}$} forces the prediction of an observed interaction to be higher than its unobserved counterparts, and the teacher score {\small $\widehat{y}^{\,tch}_{u,i}$} is computed as {\small $\widehat{y}^{\,tch}_{u,i} = \big<\big|\big|_{l=0}^L w_l\boldsymbol{v}_u^{(l)}, \big|\big|_{l=0}^L w_l\boldsymbol{v}_i^{(l)}\big>$}.
Please notice that we only disable binarization and its associated gradient estimation in pre-training.
After it is well-trained, for each user $u$, we retrieve the layer-wise teacher inference towards all items $\mathcal{I}$:
\begin{small}
\begin{equation}
\label{eq:tch_emb}
\setlength\abovedisplayskip{2pt}
\setlength\belowdisplayskip{-2pt}
\widehat{\emb{y}}^{\,tch, (l)}_{u} = \big< w_l\widehat{\emb{v}}_u^{(l)},  w_l\widehat{\emb{v}}_i^{(l)}\big>_{i \in \mathcal{I}}.
\end{equation}
\end{small}%
Based on the segment scores {\small $\boldsymbol{\widehat{y}}_u^{\,tch, (l)}$} at the $l$-th layer, we first sort out Top-R items with the highest matching scores, denoted by {\small $S^{(l)}_{tch}(u)$}. And hyper-parameter R $\ll$ $\mathcal{I}$.
Inspired by~\cite{tang2018ranking}, then we propose our layer-wise inference distillation as follows:
\begin{small}
\begin{equation}
\label{eq:tch_layer_wise}
\setlength\abovedisplayskip{2pt}
\setlength\belowdisplayskip{2pt}
\resizebox{0.9\linewidth}{!}{$
\displaystyle
\mathcal{L}_{ID}(u) = \sum_{l=0}^L \mathcal{L}_{ID}^{(l)}(\widehat{\boldsymbol{y}}_u^{\,std, (l)}, S^{(l)}_{tch}(u)) = -\frac{1}{R} \sum_{l=0}^L \sum^{R}_{k=1} w_k \cdot \ln\sigma(\widehat{y}_{u,S^{(l)}_{tch}(u, k)}^{\,std, (l)}),
$}
\end{equation}
\end{small}%
where student scores {\small $\widehat{\emb{y}}^{\,std, (l)}_{u}$} is computed similarly to Equation~(\ref{eq:tch_emb}) and {\small $S^{(l)}_{tch}(u, k)$} returns the $k$-th high-scored item from the pseudo-positive samples.
{\small $w_k$} is the ranking-aware weight presenting two major effects:
(1) since samples in {\small $S^{(l)}_{tch}(u)$} are not necessarily all ground-truth positive, $w_k$ thus balances their contribution to the overall loss;
(2) it dynamically adjusts the weight importance for different ranking positions in {\small $S^{(l)}_{tch}(u)$}.
To achieve these, $w_k$ can be developed by following the parameterized geometric distribution for approximating the tailed item popularity~\cite{rendle2014improving}:
\begin{small}
\begin{equation}
\setlength\abovedisplayskip{2pt}
\setlength\belowdisplayskip{2pt}
\label{eq:wk}
w_k = \lambda_1 \exp(-\lambda_2\cdot k),
\end{equation}
\end{small}%
where {\small $\lambda_1$} and {\small $\lambda_2$} control the loss contribution level and sharpness of the distribution. 
Intuitively, {\small $\mathcal{L}_{ID}$} encourages highly-recommended items from full-precision embeddings to more frequently appear in the student's inference list.
Moreover, our distillation approach regularizes the embedding quantization in a layer-wise manner as well;
this will effectively narrow their inference discrepancy for a more correlated recommendation capability.

\textbf{Objective Function.}
Combining {\small $\mathcal{L}^{std}_{BPR}$} that calculates BPR loss (similar to Equation~(\ref{eq:hd-bpr})) with the student predictions from Equation~(\ref{eq:score}) and {\small $\mathcal{L}_{ID}$} for all training samples, our final objective function for learning embedding binarization is defined as:
\begin{small}
\begin{equation}
\setlength\abovedisplayskip{2pt}
\setlength\belowdisplayskip{2pt}
\mathcal{L} = \mathcal{L}^{std}_{BPR} + \mathcal{L}_{ID} + \lambda ||\Theta||_2^2, 
\end{equation}
\end{small}%
where {\small $||\Theta||_2^2$} is the $L$2-regularizer of node embeddings parameterized by hyper-parameter {\small $\lambda$} to avoid over-fitting.

\vspace{-0.1in}
\subsection{Gradient Estimation}
\label{sec:gradient}
Although \textit{Straight-Through Estimator (STE)}~\cite{bengio2013estimating} enables an executable gradient flow for backpropagation, it however may cause the issue of inconsistent optimization direction in forward and backward propagation: as the integral of the constant 1 in STE is a linear function, other than $\sign(\cdot)$ function. 
To furnish more accurate gradient estimation, in this paper, we utilize the approximation of \textit{Dirac delta function}~\cite{bracewell1986fourier} for gradient estimation.

\begin{figure}[tp]
\begin{minipage}{0.5\textwidth}
\hspace{-0.2in}
\includegraphics[width=3.6in]{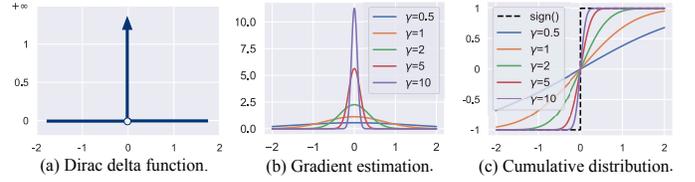}
\end{minipage} 
\setlength{\abovecaptionskip}{0.2cm}
\setlength{\belowcaptionskip}{0.2cm}
\vspace{-0.05in}
\caption{Gradient estimation.}
\label{fig:gradient}
\end{figure}

Concretely, let {\small $u(\phi)$} denote the \textit{unit-step function}, a.k.a., Heaviside step function~\cite{step_function}, where {\small $u(\phi)$ $=$ $1$} for {\small $\phi$ $>$ $0$} and {\small $u(\phi)$ $=$ $0$} otherwise. 
Obviously, we can take a translation from step function to $\sign(\cdot)$ by {\small $\sign(\phi)$ $=$ 2$u(\phi)$ - 1}, and thus theoretically {\small $\frac{\partial \sign(\phi)}{\partial \phi}$ $=$ $2\frac{\partial u(\phi)}{\partial \phi}$}. As for {\small $\frac{\partial u(\phi)}{\partial \phi}$}, it has been proved~\cite{bracewell1986fourier} that, {\small $\frac{\partial u(\phi)}{\partial \phi}$ $=$ $0$ if $\phi$ $\neq$ $0$}, and {\small $\frac{\partial u(\phi)}{\partial \phi}$ $=$ $\infty$} otherwise, which exactly is the \textit{Dirac delta function}, a.k.a., the unit impulse function {\small $\delta(\cdot)$}~\cite{bracewell1986fourier} shown in Figure~\ref{fig:gradient}(a).
However, it is still intractable to directly use {\small $\delta(\cdot)$} for gradient estimation. 
A feasible solution is to approximate {\small $\delta(\cdot)$} by introducing zero-centered Gaussian probability density as follows:
\begin{small}
\begin{equation}
\setlength\abovedisplayskip{2pt}
\setlength\belowdisplayskip{2pt}
 \delta(\phi) = \lim_{\beta \rightarrow \infty} \frac{|\beta|}{\sqrt{\pi}} \exp(-(\beta\phi)^2),
\end{equation}
\end{small}%
which imlies that: 
\begin{small}
\begin{equation}
\label{eq:norm_gradient}
\setlength\abovedisplayskip{2pt}
\setlength\belowdisplayskip{2pt}
\frac{\partial \sign(\phi)}{\partial \phi} \doteq \frac{2\gamma}{\sqrt{\pi}} \exp(-(\gamma\phi)^2).
\end{equation}
\end{small}%
As shown in Figure~\ref{fig:gradient}(b)-(c), hyper-parameter {\small $\gamma$} determines the sharpness of the derivative curve for approximation to $\sign(\cdot)$.

Intuitively, our proposed gradient estimator follows the main direction of factual gradients with $\sign(\cdot)$ in model optimization. 
This will produce a coordinated value quantization from continuous embeddings to quantized ones, and thus a series of evolving gradients can be estimated for the inputs with diverse value ranges. 
As we will show in~\cref{exp:gradient} of experiments, our gradient estimator can work better than other recent estimators~\cite{gong2019differentiable,qin2020forward,darabi2018bnn,sigmoid,RBCN}.

\section{Model Analysis}
\label{sec:analysis}

\subsection{Magnification of Feature Uniqueness}
\label{sec:necessity}
We take user $u$ as an example for illustration and the following analysis can be popularized to other nodes without loss of generality.
Similar to sensitivity analysis in statistics~\cite{koh2017understanding} and influence diffusion in social networks~\cite{xu2018representation}, we measure how the latent feature of a distant node $x$ finally affects $u$'s representation segments before binarization (e.g., {\small $\boldsymbol{v}^{(l)}_u$}), supposing $x$ is a multi-hop neighbor of $u$.
We denote the \textbf{feature enrichment ratio} {\small $\mathbb{E}_{x \rightarrow u}^{(l)}$} as the L1-norm of Jacobian matrix {\small $\begin{matrix}\left[{\partial \boldsymbol{v}^{(l)}_u}/\ {\partial \boldsymbol{v}^{(0)}_{x_u}} \right]\end{matrix}$}, by detecting the absolute influence of all fluctuation in entries of {\small $\boldsymbol{v}^{(0)}_{x}$} to {\small $\boldsymbol{v}^{(l)}_u$}, i.e., {\small $\mathbb{E}_{x\rightarrow u}^{(l)}$ = $\begin{matrix}\left|\left| \left[{\partial \boldsymbol{v}^{(l)}_u} /\ {\partial \boldsymbol{v}^{(0)}_{x}} \right] \right|\right|_1\end{matrix}$}.
Focusing on a $l$-length path $h$ connected by the node sequence: {\small $x_h^l$, $x_h^{l-1}$, $\cdots$, $x_h^1$, $x_h^0$}, where {\small $x_h^l$ = $u$} and {\small $x_h^0$ = $x$}, we follow the chain rule to develop the derivative decomposition as: 
\begin{small}
\begin{equation}
\setlength\abovedisplayskip{2pt}
\setlength\belowdisplayskip{2pt}
\begin{aligned}
\frac{\partial \boldsymbol{v}^{(l)}_u}{\partial \boldsymbol{v}^{(0)}_{x}}  = \sum_{h=1}^H
\begin{matrix}
\left[\frac{\partial \boldsymbol{v}^{(l)}_{x_h^l}}{\partial \boldsymbol{v}^{(0)}_{x_h^0}} \right]_h
\end{matrix}  
& = \sum_{h=1}^H \prod_{k=l}^1
\begin{matrix} \frac{1}{\sqrt{|\mathcal{N}(x^k_h)}|} \cdot \frac{1}{\sqrt{|\mathcal{N}(x^{k-1}_h)}|} \cdot \boldsymbol{I}\end{matrix} \\
& = \begin{matrix} \sqrt{\frac{|\mathcal{N}(u)|}{|\mathcal{N}(x)|}} \end{matrix} \sum_{h=1}^H \prod_{k=1}^l \begin{matrix} \frac{1}{|\mathcal{N}(x^k_h)|}  \cdot \boldsymbol{I} \end{matrix},\\
\end{aligned}
\end{equation}
\end{small}%
where $H$ is the number of paths between $u$ and $x$ in total.
Since all factors in the computation chain are positive, then:
\begin{small}
\begin{equation}
\setlength\abovedisplayskip{2pt}
\setlength\belowdisplayskip{2pt}
\begin{matrix}
\mathbb{E}_{x\rightarrow u}^{(l)} = \left|\left| \left[\frac{\partial \boldsymbol{v}^{(l)}_u}{\partial \boldsymbol{v}^{(0)}_{x}} \right]\right|\right|_1 = d \cdot \sqrt{\frac{|\mathcal{N}(u)|}{|\mathcal{N}(x)|}}\end{matrix}  \cdot \sum_{h=1}^H \prod_{k=1}^l \begin{matrix} \frac{1}{|\mathcal{N}(x^k_h)|}\end{matrix}.
\end{equation}
\end{small}%
Note that here the term {\small $\sum_{h=1}^H \prod_{k=1}^l 1/|\mathcal{N}(x^k_h)|$} is exactly the probability of the $l$-length random walk starting at $u$ that finally arrives at $x$, which can be interpreted as:
\begin{small}
\begin{equation}
\label{eq:prob}
\setlength\abovedisplayskip{2pt}
\setlength\belowdisplayskip{2pt}
\mathbb{E}_{x\rightarrow u}^{(l)} \propto \frac{1}{\sqrt{|\mathcal{N}(x)|}} \cdot Prob(\text{$l$-step random walk from $u$ arrives at $x$}).
\end{equation}
\end{small}%

\textbf{Magnification Effect of Feature Uniqueness.}
Equation~(\ref{eq:prob}) implies that, with the equal probability to visit adjacent neighbors, {distant nodes with fewer degrees (i.e., {\small $|\mathcal{N}(x)|$}) will contribute more feature influence to user $u$.
But most importantly, in practice, these $l$-hop neighbors of user $u$ usually represent certain \textit{esoteric} and \textit{unique} objects with less popularity.
By collecting the intermediate information in different depth of the graph convolution, we can achieve the \textbf{feature magnification effect} for all unique nodes that live within $L$ hops of graph exploration, which finally enrich $u$'s semantics in all embedding segments for quantization.


\subsection{Complexity Analysis}
\label{sec:complexity}
To discuss the feasibility for realistic deployment, we compare BiGeaR with the best full-precision model LightGCN~\cite{lightgcn}, as they are \textit{end-to-end} with offline model training and online prediction. 

\textbf{Training Time Complexity.}
{\small $M$}, {\small $N$}, and {\small $E$} represent the number of users, items, and edges; {\small $S$} and {\small $B$} are the epoch number and batch size.
We use BiGeaR$_{tch}$ and BiGeaR$_{std}$ to denote the pre-training version and binarized one, respectively.
As we can observe from Table~\ref{tab:time}, 
(1) both BiGeaR$_{tch}$ and BiGeaR$_{std}$ takes asymptotically similar complexity of graph convolution with LightGCN (where BiGeaR$_{std}$ takes additional {\small $O(2Sd(L+1)E)$} complexity for binarization).
(2) For {\small $\mathcal{L}_{BPR}$} computation, to prevent \textit{over-smoothing} issue~\cite{li2019deepgcns,li2018deeper}, usually {\small $L\leq 4$}; compare to the convolution operation, the complexity of {\small $\mathcal{L}_{BPR}$} is acceptable. 
(3) For {\small $\mathcal{L}_{ID}$} preparation, after the training of BiGeaR$_{tch}$, we firstly obtain the layer-wise prediction scores with {\small $O(MNd(L+1))$} time complexity and then sort out the Top-R pseudo-positive samples for each user with {\small $O(N + R\ln R)$}.
For BiGeaR$_{std}$, it takes a layer-wise inference distillation from BiGeaR$_{tch}$ with {\small $O\big(SRd(L+1)E \big)$}.
(4) To estimate the gradients for BiGeaR$_{std}$, it takes {\small $O(2Sd(L+1)E)$} for all training samples. 

 \begin{table}[tbh]
\setlength{\abovecaptionskip}{0.2cm}
\setlength{\belowcaptionskip}{0.2cm}
\centering
\footnotesize
\caption{Traing time complexity.}
\vspace{-0.05in}
\label{tab:time}
\setlength{\tabcolsep}{0.1mm}{
\begin{tabular}{c | c | c | c}
\toprule
  ~      & {\footnotesize LightGCN}     &  {\footnotesize BiGeaR$_{tch}$} 		&  {\footnotesize BiGeaR$_{std}$}\\
\midrule
\midrule
{\scriptsize Graph Normalization}           &   {\scriptsize $O\big(2E\big)$}           & {\scriptsize $O\big(2E\big)$}  	& {\scriptsize -}\\
\midrule[0.1pt]
{\scriptsize Conv. and Quant.}   &   {\scriptsize $O\big(\frac{2SdE^2L}{B}\big)$}     & {\scriptsize $O\big(\frac{2SdE^2L}{B}\big)$}    & {\scriptsize $O\big(2Sd(\frac{E^2L}{B}+(L+1)E)\big)$} \\
\midrule[0.1pt]
{\scriptsize $\mathcal{L}_{BPR}$ Loss}    &    {\scriptsize $O\big(2SdE\big)$}      & {\scriptsize $O\big(2Sd(L+1)E\big)$}    & {\scriptsize $O\big(2Sd(L+1)E\big)$} \\
\midrule[0.1pt]
{\scriptsize $\mathcal{L}_{ID}$ Loss}    &    {\scriptsize -}      & {\scriptsize $O\big(MNd(L+1)(N+R\ln R)\big)$}    & {\scriptsize $O\big(SRd(L+1)E \big)$} \\
\midrule[0.1pt]
{\scriptsize Gradient Estimation} &   {\scriptsize -}       & {\scriptsize -}  & {\scriptsize $O\big(2Sd(L+1)E\big)$}\\
\bottomrule
\end{tabular}}
\end{table}

\vspace*{0.3\baselineskip} 
\textbf{Memory overhead and Prediction Acceleration.}
We measure the memory footprint of embedding tables for online prediction.
As we can observe from the results in Table~\ref{tab:prediction}:
(1) Theoretically, the embedding size ratio of our model over LightGCN is {\small $\frac{32d}{(L+1)(32+d)}$}. Normally, {\small $L\leq 4$} and {\small $d\geq 64$}, thus our model achieves at least 4$\times$ space cost compression.
(2) As for the prediction time cost, we compare the number of binary operations (\#BOP) and floating-point operations (\#FLOP) between our model and LightGCN.
We find that BiGeaR replaces most of floating-point arithmetics (e.g., multiplication) with bitwise operations. 
\begin{table}[tbh]
\setlength{\abovecaptionskip}{0.2cm}
\setlength{\belowcaptionskip}{0.2cm}
\centering
\footnotesize
\caption{Complexity of space cost and online prediction.}
\vspace{-0.05in}
\label{tab:prediction}
\setlength{\tabcolsep}{2.2mm}{
\begin{tabular}{c | c | c c}
\toprule
  ~          & {\footnotesize Embedding size}   &  {\footnotesize \#FLOP}	& {\footnotesize \#BOP}     	\\
\midrule
\midrule
 {\scriptsize LightGCN}      & {\scriptsize $O\big(32(M+N)d\big)$}       &   {\scriptsize $O\big(2MNd\big)$}      &   {-}       	\\
\midrule[0.1pt]
{\scriptsize BiGeaR}       & {\scriptsize $O\big((M+N)(L+1)(32+d)\big)$}     & {\scriptsize $O\big(4MN(L+1)\big)$}            & {\scriptsize $O\big(2MN(L+1)d\big)$}   	\\
\bottomrule
\end{tabular}}
\end{table}

\section{Experimental Results}
\label{sec:exp}
We evaluate our model on Top-K recommendation task with the aim of answering the following research questions:
\begin{itemize}[leftmargin=*,topsep=0.5pt,parsep=0.5pt]
\item \textbf{RQ1.} How does BiGeaR perform compared to state-of-the-art full-precision and quantization-based models?

\item \textbf{RQ2.} How is the practical resource consumption of BiGeaR?

\item \textbf{RQ3.} How do proposed components affect the performance?

\end{itemize} 

\subsection{Experiment Setup}

\noindent\textbf{Datasets.}
To guarantee the fair comparison, we directly use five widely evaluated datasets (including the training/test splits) from: MovieLens\footnote{\url{https://grouplens.org/datasets/movielens/1m/}}~\cite{hashgnn,he2016fast,chen2021modeling,chen2021attentive}, Gowalla\footnote{\url{https://github.com/gusye1234/LightGCN-PyTorch/tree/master/data/gowalla}}~\cite{ngcf,hashgnn,lightgcn,dgcf}, Pinterest\footnote{\url{https://sites.google.com/site/xueatalphabeta/dataset-1/pinterest_iccv}}~\cite{geng2015learning,hashgnn}, Yelp2018\footnote{\url{https://github.com/gusye1234/LightGCN-PyTorch/tree/master/data/yelp2018}}~\cite{ngcf,dgcf,lightgcn}, Amazon-Book\footnote{\url{https://github.com/gusye1234/LightGCN-PyTorch/tree/master/data/amazon-book}}~\cite{ngcf,dgcf,lightgcn}.
Dataset statistics and descriptions are reported in Table~\ref{tab:datasets} and Appendix~\ref{app:dataset}.

\begin{table}[h]
\setlength{\abovecaptionskip}{0.2cm}
\setlength{\belowcaptionskip}{0.2cm}
\centering
\small
\caption{The statistics of datasets.}
\vspace{-0.05in}
\label{tab:datasets}
\setlength{\tabcolsep}{0.7mm}{
\begin{tabular}{c | c  c  c  c  c}
\toprule 
             & { MovieLens}  & { Gowalla}   & { Pinterest}  &  { Yelp2018} & { Amazon-Book}\\
\midrule
\midrule[0.1pt]
    {\#Users}  & {6,040}   & {29,858}   & {55,186}   & {31,668}  &{52,643}  \\ 
    {\#Items}  & {3,952}   & {40,981}   & {9,916}    & {38,048}  &{91,599}  \\
\midrule[0.1pt]
    {\#Interactions} & {1,000,209} & {1,027,370} & {1,463,556} & {1,561,406} & {2,984,108} \\
\bottomrule
\end{tabular}}
\end{table}

\noindent\textbf{Evaluation Metric.}
In Top-K recommendation evaluation, we select two widely-used evaluation protocols Recall@K and NDCG@K to evaluate Top-K recommendation capability.

\vspace*{0.3\baselineskip} 
\noindent\textbf{Competing Methods.}
We include the following recommender models: 
(1) 1-bit quantization-based methods including graph-based (GumbelRec~\cite{gumbel1,gumbel2,zhang2019doc2hash}, HashGNN~\cite{hashgnn}) and general model-based (LSH~\cite{lsh}, HashNet~\cite{hashnet}, CIGAR~\cite{kang2019candidate}), 
and (2) full-precision  models including neural-network-based (NeurCF~\cite{neurcf}) and graph-based (NGCF~\cite{ngcf}, DGCF~\cite{dgcf}, LightGCN~\cite{lightgcn}).
Detailed introduction of these methods are attached in Appendix~\ref{app:method}.

We exclude early quantization-based recommendation baselines, e.g., CH~\cite{liu2014collaborative}, DiscreteCF~\cite{zhang2016discrete},  DPR~\cite{zhang2017discrete}, and full-precision solutions, e.g., GC-MC~\cite{berg2017graph}, PinSage~\cite{pinsage}, mainly because the above competing models~\cite{kang2019candidate,lightgcn,ngcf,neurcf} have validated the superiority.

\vspace*{0.3\baselineskip} 

\noindent\textbf{Experiment Settings.}
Our model is implemented by Python 3.7 and PyTorch 1.14.0 with non-distributed training. 
The experiments are run on a Linux machine with 1 NVIDIA V100 GPU, 4 Intel Core i7-8700 CPUs, 32 GB of RAM with 3.20GHz.
For all the baselines, we follow the official reported hyper-parameter settings, and for methods lacking recommended settings, we apply a grid search for hyper-parameters.
The embedding dimension is searched in \{$32, 64, 128, 256, 512, 1024$\}. 
The learning rate $\eta$ is tuned within \{$10^{-4}, 10^{-3}, 10^{-2}$\} and the coefficient of $L2$ normalization $\lambda$ is tuned among \{$10^{-6}, 10^{-5}, 10^{-4}, 10^{-3}$\}. 
We initialize and optimize all models with default normal initializer and Adam optimizer~\cite{adam}. 
We report all hyper-parameters in Appendix~\ref{app:parameter} for reproducibility.

\begin{table*}[tbh]
\setlength{\abovecaptionskip}{0.2cm}
\setlength{\belowcaptionskip}{0.2cm}
\centering
\small
  \caption{\small Performance comparison (waveline and underline represent the best performing full-precision and quantization-based models). }
  \vspace{-0.05in}
  \label{tab:top20}
  \setlength{\tabcolsep}{1.4mm}{
  \begin{tabular}{c|c c|c c|c c|c c|c c} 
    \toprule
    \multirow{2}*{Model} & \multicolumn{2}{c|}{MovieLens (\%)} & \multicolumn{2}{c|}{Gowalla (\%)} & \multicolumn{2}{c|}{Pinterest (\%)} & \multicolumn{2}{c|}{Yelp2018 (\%)} &  \multicolumn{2}{c}{Amazon-Book (\%)} \\
        ~ & Recall@20 & NDCG@20  & Recall@20 & NDCG@20  & Recall@20  & NDCG@20  & Recall@20  & NDCG@20  & Recall@20  & NDCG@20 \\
    \midrule
    \midrule
    NeurCF           & {21.40} $\pm$ {\footnotesize 1.51}   & {37.91} $\pm$ {\footnotesize 1.14}   & {14.64} $\pm$ {\footnotesize 1.75}  & {23.17} $\pm$ {\footnotesize 1.52} & {12.28} $\pm$ {\footnotesize 1.88} & {13.41} $\pm$ {\footnotesize 1.13}   & {4.28} $\pm$ {\footnotesize 0.71}   & {7.24} $\pm$ {\footnotesize 0.53}  &{3.49} $\pm$ {\footnotesize 0.75}  &{6.71} $\pm$ {\footnotesize 0.72} \\
    NGCF             & {24.69} $\pm$ {\footnotesize 1.67}  & {39.56} $\pm$ {\footnotesize 1.26}   & {16.22} $\pm$ {\footnotesize 0.90}  & {24.18} $\pm$ {\footnotesize 1.23}  & {14.67} $\pm$ {\footnotesize 0.56}  & {13.92} $\pm$ {\footnotesize 0.44} & {5.89} $\pm$ {\footnotesize 0.35} & {9.38} $\pm$ {\footnotesize 0.52}   &{3.65} $\pm$ {\footnotesize 0.73} &{6.90} $\pm$ {\footnotesize 0.65} \\
    DGCF             & {25.28} $\pm$ {\footnotesize 0.39}  & {45.98} $\pm$ {\footnotesize 0.58}   & {18.64} $\pm$ {\footnotesize 0.30}  & {25.20} $\pm$ {\footnotesize 0.41} & {\uwave{15.52}} $\pm$ {\footnotesize 0.42} & {\uwave{16.51}} $\pm$ {\footnotesize 0.56} & {6.37} $\pm$ {\footnotesize 0.55}  & {11.08} $\pm$ {\footnotesize 0.48}   &{4.32} $\pm$ {\footnotesize 0.34} &{7.73} $\pm$ {\footnotesize 0.27}\\
    LightGCN          & {\uwave{26.28}}  $\pm$ {\footnotesize 0.20}  & {\uwave{46.04}} $\pm$ {\footnotesize 0.18}   & {\uwave{19.02}}  $\pm$ {\footnotesize 0.19}  & {\uwave{25.71}} $\pm$ {\footnotesize 0.25}  & {15.33} $\pm$ {\footnotesize 0.28}  & {16.29} $\pm$ {\footnotesize 0.24}  & {\uwave{6.79}} $\pm$ {\footnotesize 0.31}  & {\uwave{12.17}} $\pm$ {\footnotesize 0.27} &{\uwave{4.84}} $\pm$ {\footnotesize 0.09}  &{\uwave{8.11}} $\pm$ {\footnotesize 0.11} \\
    \midrule[0.1pt]
    \cellcolor{best}\textbf{BiGeaR}  &{\textbf{25.57}} $\pm$ {\footnotesize 0.33} &{\textbf{45.56}} $\pm$ {\footnotesize 0.31} &{\textbf{18.36}} $\pm$ {\footnotesize 0.14} &{\textbf{24.96}} $\pm$ {\footnotesize 0.17} &{\textbf{15.57}} $\pm$ {\footnotesize 0.22} &{\textbf{16.83}} $\pm$ {\footnotesize 0.46} &{\textbf{6.47}} $\pm$ {\footnotesize 0.14} &{\textbf{11.60}} $\pm$ {\footnotesize 0.18} &{\textbf{4.68}} $\pm$ {\footnotesize 0.11} &{\textbf{8.12}} $\pm$ {\footnotesize 0.12} \\   
    \textbf{Capability} &{97.30\%} &{98.96\%} &{96.53\%} &{97.08\%} &{100.32\%} &{101.94\%} &{95.29\%} &{95.32\%} &{96.69\%} &{100.12\%} \\
    \midrule 
    LSH             & {11.38} $\pm$ {\footnotesize 1.23}  & {14.87} $\pm$ {\footnotesize 0.76}   & {8.14} $\pm$ {\footnotesize 0.98}   & {12.19} $\pm$ {\footnotesize 0.86} & {7.88} $\pm$ {\footnotesize 1.21}  & {9.84} $\pm$ {\footnotesize 0.90}   & {2.91} $\pm$ {\footnotesize 0.51}   & {5.06} $\pm$ {\footnotesize 0.67} &{2.41}	$\pm$ {\footnotesize 0.95} &{4.39}	$\pm$ {\footnotesize 1.16} \\
    HashNet         & {15.43} $\pm$ {\footnotesize 1.73} & {24.78} $\pm$ {\footnotesize 1.50}   & {11.38} $\pm$ {\footnotesize 1.25}   & {16.50} $\pm$ {\footnotesize 1.42} & {10.27}  $\pm$ {\footnotesize 1.48} & {11.64} $\pm$ {\footnotesize 0.91}  & {3.37}  $\pm$ {\footnotesize 0.78}  & {7.31} $\pm$ {\footnotesize 1.16}  &{2.86}	$\pm$ {\footnotesize 1.51}	&{4.75}	$\pm$ {\footnotesize 1.33} \\
    CIGAR & {14.84} $\pm$ {\footnotesize 1.44} & {24.63} $\pm$ {\footnotesize 1.77}   & {11.57} $\pm$ {\footnotesize 1.01}   & {16.77} $\pm$ {\footnotesize 1.29} & {10.34}  $\pm$ {\footnotesize 0.97} & {11.87} $\pm$ {\footnotesize 1.20}  & {3.65}  $\pm$ {\footnotesize 0.90}  & {7.87} $\pm$ {\footnotesize 1.03}  &{3.05}  $\pm$ {\footnotesize 1.32}  &{4.98} $\pm$ {\footnotesize 1.24} \\
    GumbelRec       & {16.62} $\pm$ {\footnotesize 2.17} & {29.36} $\pm$ {\footnotesize 2.53} & {12.26}  $\pm$ {\footnotesize 1.58} & {17.49} $\pm$ {\footnotesize 1.08} & {10.53} $\pm$ {\footnotesize 0.79} & {11.86} $\pm$ {\footnotesize 0.86} & {3.85} $\pm$ {\footnotesize 1.39} & {7.97} $\pm$ {\footnotesize 1.59} &{2.69} $\pm$ {\footnotesize 0.55} &{4.32}  $\pm$ {\footnotesize 0.47} \\
    HashGNN$_h$     & {14.21} $\pm$ {\footnotesize 1.67} & {24.39} $\pm$ {\footnotesize 1.87}  & {11.63} $\pm$ {\footnotesize 1.47}  & {16.82} $\pm$ {\footnotesize 1.35}  & {10.15} $\pm$ {\footnotesize 1.43} & {11.96}  $\pm$ {\footnotesize 1.58} & {3.77}  $\pm$ {\footnotesize 1.02} & {7.75} $\pm$ {\footnotesize 1.39} &{3.09}	$\pm$ {\footnotesize 1.29} &{5.19} $\pm$ {\footnotesize 1.03}	\\
    HashGNN$_s$     & {\underline{19.87}} $\pm$ {\footnotesize 0.93} & {\underline{37.32}} $\pm$ {\footnotesize 0.81}  & {\underline{13.45}} $\pm$ {\footnotesize 0.65}  & {\underline{19.12}} $\pm$ {\footnotesize 0.68} & {\underline{12.38}}  $\pm$ {\footnotesize 0.86} & {\underline{13.63}} $\pm$ {\footnotesize 0.75} & {\underline{4.86}} $\pm$ {\footnotesize 0.36} & {\underline{8.83}} $\pm$ {\footnotesize 0.27} &{\underline{3.34}} $\pm$ {\footnotesize 0.25} 	&{\underline{5.82}} $\pm$ {\footnotesize 0.24}	\\
    \midrule[0.1pt]
    \cellcolor{best}\textbf{BiGeaR}  &{\textbf{25.57}} $\pm$ {\footnotesize 0.33} &{\textbf{45.56}} $\pm$ {\footnotesize 0.31} &{\textbf{18.36}} $\pm$ {\footnotesize 0.14} &{\textbf{24.96}} $\pm$ {\footnotesize 0.17} &{\textbf{15.57}} $\pm$ {\footnotesize 0.22} &{\textbf{16.83}} $\pm$ {\footnotesize 0.46} &{\textbf{6.47}} $\pm$ {\footnotesize 0.14} &{\textbf{11.60}} $\pm$ {\footnotesize 0.18} &{\textbf{4.68}} $\pm$ {\footnotesize 0.11} &{\textbf{8.12}} $\pm$ {\footnotesize 0.12} \\   
    \textbf{Gain}   &{28.69\%} &{22.08\%} &{36.51\%} &{30.54\%} &{25.77\%} &{23.48\%} &{33.13\%} &{31.37\%}  &{40.12\%} &{39.52\%} \\
    \textbf{$p$-value} &{5.57e-7} &{2.64e-8} &{2.21e-7} &{7.69e-8} &{2.5e-5} &{3.51e-5} &{3.27e-6} &{5.30e-8} &{3.49e-6} &{7.14e-8} \\
    \bottomrule
  \end{tabular}}
\end{table*}

\subsection{Performance Analysis (RQ1)}
We evaluate Top-K recommendation by varying K in \{20, 40, 60, 80, 100\}.
We summarize the Top@20 results in Table~\ref{tab:top20} for detailed comparison and plot the Top-K recommendation curves in Appendix~\ref{app:topk}. From Table~\ref{tab:top20}, we have the following observations:
\begin{itemize}[leftmargin=*,topsep=0.5pt,parsep=0.5pt]
\item \textbf{Our model offers a competitive recommendation capability to state-of-the-art full-precision recommender models.}
(1) BiGeaR generally outperforms most of full-precision recommender models excluding LightGCN over five benchmarks.
The main reason is that our model and LightGCN take similar graph convolution methodology with network simplification~\cite{lightgcn}, e.g., removing self-connection and feature transformation, which is proved to be effective for Top-K ranking and recommendation.
Moreover, BiGeaR collects the different levels of interactive information in multi depths of graph exploration, which significantly enriches semantics to user-item representations for binarization.
(2) Compared to the state-of-the-art method LightGCN, our model develops about 95\%$\sim$102\% of performance capability \textit{w.r.t.} Recall@20 and NDCG@20 throughout all datasets.
This shows that our proposed model designs are effective to narrow the performance gap with full-precision model LightGCN. 
Although the binarized embeddings learned by BiGeaR may not achieve the \textit{exact} expressivity parity with the full-precision ones learned by LightGCN, considering the advantages of space compression and inference acceleration that we will present later, we argue that such performance capability is acceptable, especially for those resource-limited deployment scenarios.

\item \textbf{Compared to all binarization-based recommender models, BiGeaR presents the empirically remarkable and statistically significant performance improvement.}
(1) Two conventional methods (LSH, HashNet) for general item retrieval tasks underperform CIGAR, HashGNN and BiGeaR, showing that a direct model adaptation may be too trivial for Top-K recommendation. 
(2) Compared to CIGAR, graph-based models generally work better.
This is mainly because, CIGAR combines general neural networks with \textit{learning to hash} techniques for fast candidate generation;
on the contrary, graph-based models are more capable of exploring multi-hop interaction subgraphs to directly simulate the high-order \textit{collaborative filtering} process for model learning.
(3) Our model further outperforms HashGNN by about 26\%$\sim$40\% and 22\%$\sim$40\% \textit{w.r.t.} Recall@20 and NDCG@20, proving the effectiveness of our proposed multi-faceted optimization components in embedding binarization. 
(4) Moreover, the significance test in which $p$-value $<$ 0.05 indicates that the improvements over all five benchmarks are statistically significant.

\end{itemize}

\subsection{Resource Consumption Analysis (RQ2)}
We analyze the resource consumption in \textit{training}, \textit{online inference}, and \textit{memory footprint} by comparing to the best two competing models, i.e., LightGCN and HashGNN.
Due to the page limits, we report the empirical results of MovieLens dataset in Table~\ref{tab:consumption}.
\begin{enumerate}[leftmargin=*,topsep=0.5pt,parsep=0.5pt]
\item $T_{train}$: we set batch size $B=2048$ and dimension size $d=256$ for all models. 
We find that HashGNN is fairly time-consuming than LightGCN and BiGeaR. 
This is because HashGNN adopts the early GCN framework~\cite{graphsage} as the backbone; LightGCN and BiGeaR utilize more simplified graph convolution architecture in which operations such as self-connection, feature transformation, and nonlinear activation are all removed~\cite{lightgcn}.
Furthermore, BiGeaR needs 5.1s and 6.2s per epoch for pretraining and quantization, both of which take slightly more yet asymptotically similar time cost with LightGCN, basically following the complexity analysis in~\cref{sec:complexity}.

\item $T_{infer}$: we randomly generate 1,000 queries for online prediction and conduct experiments with the vanilla NumPy\footnote{{\url{https://www.lfd.uci.edu/~gohlke/pythonlibs/}}} on CPUs.
We observe that HashGNN$_s$ takes a similar time cost with LightGCN.
This is because, while HashGNN$_h$ purely binarizes the continuous embeddings, its relaxed version HashGNN$_s$ adopts a Bernoulli random variable to provide the probability of replacing the quantized digits with original real values~\cite{hashgnn}.
Thus, although HashGNN$_h$ can use Hamming distance for prediction acceleration, HashGNN$_s$ with the improved recommendation accuracy can only be computed by floating-point arithmetics.
As for BiGeaR, thanks to its bitwise-operation-supported capability, it runs about 8$\times$ faster than LightGCN whilst maintaining the similar performance on MovieLens dataset.

\item $S_{ET}$: we only store the embedding tables that are necessary for online inference. 
As we just explain, HashGNN$_s$ interprets embeddings by randomly selected real values, which, however, leads to the expansion of space consumption. 
In contrast to HashGNN$_s$, BiGeaR can separately store the binarized embeddings and corresponding scalers, making a balanced trade-off between recommendation accuracy and space overhead.
\end{enumerate}

\begin{table}[th]
\setlength{\abovecaptionskip}{0.2cm}
\setlength{\belowcaptionskip}{0.2cm}
\centering
\footnotesize
\caption{Resource consumption on MovieLens dataset.}
\vspace{-0.05in}
\label{tab:consumption}
\setlength{\tabcolsep}{2.5mm}{
\begin{tabular}{c | c  c  c  c}
\toprule
              & LightGCN   &  HashGNN$_h$ &  HashGNN$_s$  &  BiGeaR\\
\midrule[0.1pt]
\midrule[0.1pt]
  $T_{train}\mathbin{/}${\scriptsize \#epcoch}     &   {4.91s}     &   {186.23s}  & {204.53s}    &   {(5.16+6.22)s}\\
\midrule[0.1pt]
   $T_{infer}\mathbin{/}${\scriptsize \#query}    & {32.54ms}    & {2.45ms}  & {31.76ms}   & {3.94ms}\\
\midrule[0.1pt]
  $S_{ET}$    & {9.79MB}    & {0.34MB}  & {9.78MB}   & {1.08MB}\\
\midrule
\midrule
  Recall@20    & {26.28\%}    & {14.21\%}  & {19.87\%}   & {25.57\%}\\
\bottomrule
\end{tabular}}
\end{table}

\begin{figure}[tp]
\begin{minipage}{0.5\textwidth}
\hspace{-0.1in}
\includegraphics[width=3.5in]{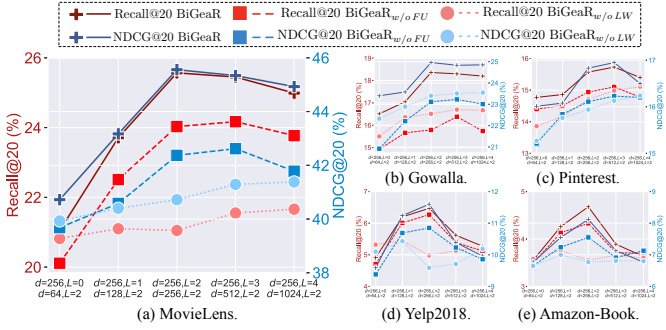}
\end{minipage} 
\setlength{\abovecaptionskip}{0.2cm}
\setlength{\belowcaptionskip}{0.2cm}
\vspace{-0.05in}
\caption{Study of graph layer-wise quantization. }
\label{fig:dim_layer}
\end{figure}
 
\subsection{Study of Layer-wise Quantization (RQ3.A)}
\label{sec:lw_exp}
To verify the magnification of feature uniqueness in layer-wise quantization, we modify BiGeaR and propose two variants, denoted as BiGeaR$_{w/o\,LW}$ and BiGeaR$_{w/o\,FU}$. 
We report the results in Figure~\ref{fig:dim_layer} by denoting Recall@20 and NDCG@20 in red and blue, respectively, and vary color brightness for different variants. From these results, we have the following explanations.
\begin{itemize}[leftmargin=*,topsep=0.5pt,parsep=0.5pt]
\item Firstly, BiGeaR$_{w/o\,LW}$ discards the layer-wise quantization and adopts the conventional manner by quantizing the last outputs from $L$ convolution iterations. 
We fix dimension $d=256$ and vary layer number $L$ for BiGeaR, and only vary dimension $d$ for variant BiGeaR$_{w/o\,LW}$ with fixed $L=2$.
(1) Even by continuously increasing the dimension size from 64 to 1024, BiGeaR$_{w/o\,LW}$ slowly improves both Recall@20 and NDCG@20 performance. 
(2) By contrast, our layer-wise quantization presents a more efficient capability in improving performance by increasing $L$ from 0 to 3.
When $L=4$, BiGeaR usually exhibits a conspicuous performance decay, mainly because of the common \textit{over-smoothing issue} in graph-based models~\cite{li2019deepgcns,li2018deeper}.
Thus, with a moderate size $d=256$ and convolution number $L \leq 3$, BiGeaR can attain better performance with acceptable computational complexity.

\item Secondly, BiGeaR$_{w/o\,FU}$ omits the feature magnification effect by adopting the way used in HashGNN~\cite{graphsage,hashgnn} as:
\begin{equation}
\setlength\abovedisplayskip{0pt}
\setlength\belowdisplayskip{0pt}
\begin{matrix}
\boldsymbol{v}^{(l)}_x = \sum_{z\in \mathcal{N}(x)} \frac{1}{|\mathcal{N}(z)|}\boldsymbol{v}^{(l-1)}_z.
\end{matrix}
\end{equation}
Similar to the analysis in~\cref{sec:necessity}, such modification will finally disable the ``magnification term'' in Equation~(\ref{eq:prob}) and simplify it to the vanilla random walk for graph exploration. 
Although BiGeaR$_{w/o\,FU}$ presents similar curve trends with BiGeaR when $L$ increases, the general performance throughout all five datasets is unsatisfactory compared to BiGeaR.
This validates the effectiveness of BiGeaR's effort in magnifying unique latent features, which enriches user-item representations and boosts Top-K recommendation performance accordingly.
\end{itemize}

\subsection{Study of Inference Distillation (RQ3.B)} 

\subsubsection{\textbf{Effect of Layer-wise Distillation.}}
We study the effectiveness of our inference distillation by proposing two ablation variants, namely \textsl{noID} and \textsl{endID}.
While \textsl{noID} totally removes our information distillation in model training, \textsl{endID} downgrades the original layer-wise distillation to only distill information from the last layer of graph convolution. 
As shown in Table~\ref{tab:distillation}, both \textsl{noID} and \textsl{endID} draw notable performance degradation.
Furthermore, the performance gap between \textsl{endID} and BiGeaR shows that it is efficacious to conduct our inference distillation in a layer-wise manner for further performance enhancement.

\begin{table}[h]
\setlength{\abovecaptionskip}{0.2cm}
\setlength{\belowcaptionskip}{0.2cm}
\centering
\footnotesize
\caption{Learning inference distillation.}
\vspace{-0.05in}
\label{tab:distillation}
\setlength{\tabcolsep}{0.6mm}{
\begin{tabular}{c |c c|c c|c c|c c|c c}
\toprule
 \multirow{2}*{Variant} & \multicolumn{2}{c|}{MovieLens} & \multicolumn{2}{c|}{Gowalla} & \multicolumn{2}{c|}{Pinterest} & \multicolumn{2}{c|}{Yelp2018} & \multicolumn{2}{c}{Amazon-book} \\
               ~  & R@20 & N@20 & R@20 & N@20 & R@20 & N@20 & R@20 & N@20 & R@20 & N@20\\
\midrule
\midrule
    \multirow{2}*{\footnotesize \textsl{noID}}    &{24.40}& {44.06}   & {17.85}& {24.28}  & {15.23}& {16.38}  & {6.18} & {11.22} & {4.07} & {7.31}\\
~        &\textit{\color{blue} \scriptsize{-4.58\%}}  &\textit{\color{blue} \scriptsize{-3.29\%}}  &\textit{\color{blue} \scriptsize{-2.78\%}}  &\textit{\color{blue} \scriptsize{-2.72\%}}  &\textit{\color{blue} \scriptsize{-2.18\%}}  &\textit{\color{blue} \scriptsize{-2.85\%}}  &\textit{\color{blue} \scriptsize{-4.48\%}}  &\textit{\color{blue} \scriptsize{-3.28\%}} &\textit{\color{blue} \scriptsize{-13.03\%}}  &\textit{\color{blue}  \scriptsize{-9.98\%}}\\
  \midrule[0.1pt]
    \multirow{2}*{\footnotesize \textsl{endID}}    &{25.02}& {44.75}   & {18.05}& {24.73}  & {15.28}& {16.58}  & {6.29} & {11.37} & {4.44} & {7.78}\\
~        &\textit{\color{blue} \scriptsize{-2.15\%}}  &\textit{\color{blue} \scriptsize{-1.78\%}}  &\textit{\color{blue} \scriptsize{-1.69\%}}  &\textit{\color{blue} \scriptsize{-0.92\%}}  &\textit{\color{blue} \scriptsize{-1.86\%}}  &\textit{\color{blue} \scriptsize{-1.49\%}}  &\textit{\color{blue} \scriptsize{-2.78\%}}  &\textit{\color{blue} \scriptsize{-1.98\%}} &\textit{\color{blue} \scriptsize{-5.13\%}}  &\textit{\color{blue}  \scriptsize{-4.19\%}}\\
\midrule[0.1pt]
  \multirow{2}*{\footnotesize \textsl{KLD}}    &{24.32}& {44.38}   & {17.63}& {24.07}  & {14.78}& {15.92}  & {5.83} & {10.36} & {4.13} & {7.21}\\
  ~        &\textit{\color{blue} \scriptsize{-4.89\%}}  &\textit{\color{blue} \scriptsize{-2.59\%}}  &\textit{\color{blue} \scriptsize{-3.98\%}}  &\textit{\color{blue} \scriptsize{-3.57\%}}  &\textit{\color{blue} \scriptsize{-5.07\%}}  &\textit{\color{blue} \scriptsize{-5.41\%}}  &\textit{\color{blue} \scriptsize{-9.89\%}}  &\textit{\color{blue} \scriptsize{-10.69\%}} &\textit{\color{blue} \scriptsize{-11.75\%}}  &\textit{\color{blue}  \scriptsize{-11.21\%}}\\
  \midrule[0.1pt]
\cellcolor{best}{\footnotesize \textbf{BiGeaR} }  &\textbf{25.57}& \textbf{45.56}   & \textbf{18.36}& \textbf{24.96}  & \textbf{15.57}& \textbf{16.83}  & \textbf{6.47}& \textbf{11.60} & \textbf{4.68}& \textbf{8.12}\\
\bottomrule
\end{tabular}}
\end{table}

\subsubsection{\textbf{Conventional Knowledge Distillation.}}
To compare with the conventional approach, we modify BiGeaR by applying KL divergence for layer-wise teacher and student logits, i.e., {\small $\boldsymbol{\widehat{y}}_u^{\,tch, (l)}$} v.s. {\small $\boldsymbol{\widehat{y}}_u^{\,std, (l)}$}.
We denote this variant as \textsl{KLD}.
As we can observe from Table~\ref{tab:distillation}, using conventional knowledge distillation with KL divergence leads to suboptimal performance. 
This is because KL divergence encourages the teacher and student training objects to have a similar logit distribution, but users' relative order of item preference can not be well learned from this process.
Compared to the conventional approach, our proposed layer-wise Inference distillation is thus more effective for ranking information distillation.

\subsection{Study of Gradient Estimation (RQ3.C)}
\label{exp:gradient}
We compare our gradient estimation with several recently studied estimators, such as \textit{Tanh-like}~\cite{qin2020forward,gong2019differentiable}, \textit{SSwish}~\cite{darabi2018bnn}, \textit{Sigmoid}~\cite{sigmoid}, and \textit{projected-based estimator}~\cite{RBCN} (denoted as PBE), by implementing them in BiGeaR.
We report their Recall@20 in Figure~\ref{fig:quant_f} and compute the performance gain of our approach over these estimators accordingly. 
We have two main observations:
\begin{enumerate}[leftmargin=*,topsep=0.5pt,parsep=0.5pt]
\item Our proposed approach shows the consistent superiority over all other gradient estimators.
These estimators usually use \textit{visually similar} functions, e.g., tanh($\cdot$), to approximate $\sign(\cdot)$.
However, these functions are not necessarily \textit{theoretically relevant} to $\sign(\cdot)$. This may lead to inaccurate gradient estimation.
On the contrary, as we explain in~\cref{sec:gradient}, we transfer the unit-step function $u(\cdot)$ to $\sign(\cdot)$ by $\sign(\cdot)$ = 2$u(\cdot)$ - 1.
Then we can further estimate the gradients of $\sign(\cdot)$ with the approximated derivatives of $u(\cdot)$.
In other words, our approach follows the main optimization direction of factual gradients with $\sign(\cdot)$;
and different from previous solutions, this guarantees the coordination in both forward and backward propagation. 

\item Furthermore, compared to the last four datasets, MovieLens dataset confronts a larger performance disparity between our approach and others.
This is because MovieLens dataset is much denser than the other datasets, i.e., $\frac{\#Interactions}{\#Users \cdot \#Items}$ = 0.0419 $\gg$ \{0.00084, 0.00267, 0.0013, 0.00062\}, which means that users tend to have more item interactions and complicated preferences towards different items. 
Consequently, this posts a higher requirement for the gradient estimation capability in learning ranking information.
Fortunately, the empirical results in Figure~\ref{fig:quant_f} demonstrate that our solution well fulfills these requirements, especially for dense interaction graphs.
\end{enumerate}

\begin{figure}[t]
\begin{minipage}{0.5\textwidth}
\includegraphics[width=3.3in]{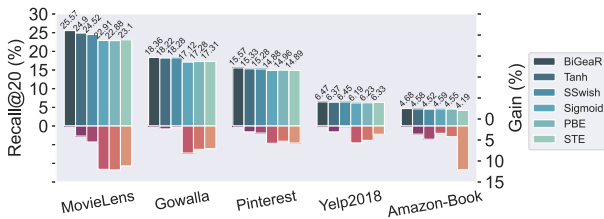}
\end{minipage} 
\setlength{\abovecaptionskip}{0.2cm}
\setlength{\belowcaptionskip}{0.2cm}
\vspace{-0.15in}
\caption{Gradient estimator comparison \textit{w.r.t.} Recall@20.}
\label{fig:quant_f}
\end{figure}

\section{Related Work}
\label{sec:work}

\textbf{Full-precision recommender models.}
(1) \textit{Collaborative Filtering (CF)} is a prevalent methodology in modern recommender systems~\cite{covington2016deep,pinsage,yang2022hrcf}. 
Earlier CF methods, e.g., \textit{Matrix Factorization}~\cite{koren2009matrix,rendle2012bpr}, reconstruct historical interactions to learn user-item embeddings.
Recent neural-network-based models, e.g., NeurCF~\cite{neurcf} and attention-based models~\cite{chen2017attentive,he2018nais}, further boost performance with neural networks.
(2) \textit{Graph-based} methods focus on studying the interaction graph topology for recommendation. 
Graph convolutional networks (GCNs)~\cite{graphsage,kipf2016semi} inspire early work, e.g., GC-MC~\cite{berg2017graph}, PinSage~\cite{pinsage}, and recent models, e.g., NGCF~\cite{ngcf}, DGCF~\cite{dgcf}, and LightGCN~\cite{lightgcn}, mainly because they can well simulate the high-order CF signals among high-hop neighbors for recommendation.


\textbf{Learning to hash.}
Hashing-based methods map dense floating-point embeddings into binary spaces for \textit{Approximate Nearest Neighbor} (ANN) search acceleration.
A representative model LSH~\cite{lsh} has inspired a series of work for various tasks, e.g., fast retrieval of images~\cite{hashnet}, documents~\cite{li2014two,zhang2020discrete}, and categorical information~\cite{kang2021learning}.
For Top-K recommendation, models like DCF~\cite{zhang2016discrete}, DPR~\cite{zhang2017discrete} include neural network architectures.
A recent work CIGAR~\cite{kang2019candidate} proposes adaptive model designs for fast candidate generation.
To investigate the graph structure of user-item interactions, model HashGNN~\cite{hashgnn} applies hashing techniques into graph neural networks for recommendation.
However, one major issue is that solely using learned binary codes for prediction usually draws a large performance decay.
Thus, to alleviate the issue, CIGAR further equips with additional full-precision recommender models (e.g., BPR-MF~\cite{rendle2012bpr}) for fine-grained \textit{re-ranking};
HashGNN proposes relaxed version by mixing full-precision and binary embedding codes.


\textbf{Quantization-based models.}
Quantization-based models share similar techniques with hashing-based methods, e.g., $\sign(\cdot)$ is usually adopted mainly because of its simplicity.
However, quantization-based models do not pursue extreme encoding compression, and thus they develop multi-bit, 2-bit, and 1-bit quantization for performance adaptation.
Recently, there is growing attention to quantize graph-based models, such as Bi-GCN~\cite{bigcn} and BGCN~\cite{bahri2021binary}
However, these two models are mainly designed for geometric classification tasks, but their capability in product recommendation is unclear.
Thus, in this paper, we propose BiGeaR to learn 1-bit user-item representation quantization for Top-K recommendation. 
Different from binary hashing-based methods, BiGeaR aims to make predictions within its own framework, making a balanced trade-off between efficiency and performance.

\section{Conclusion and Future Work}
\label{sec:con}
In this paper, we present BiGeaR to learn binarized graph representations for recommendation with multi-faceted binarization techniques.
The extensive experiments not only validate the performance superiority over competing binarization-based recommender systems, but also justify the effectiveness of all proposed model components.
In the future, we plan to investigate two major possible problems. 
(1) It is worth developing binarization techniques for model-agnostic recommender systems with diverse learning settings~\cite{song2021semi,yang2021discrete,song2022graph}. 
(2) Instead of using $\sign(\cdot)$ for quantization, developing compact multi-bit quantization methods with similarity-preserving is promising to improve ranking accuracy.


\begin{acks}
The work described in this paper was partially supported by the National Key Research and Development Program of China (No. 2018AAA0100204), the Research Grants Council of the Hong Kong Special Administrative Region, China (CUHK 2410021, Research Impact Fund, No. R5034-18), and the CUHK Direct Grant (4055147).
\end{acks}


\bibliographystyle{ACM-Reference-Format}
{
\bibliography{ref}


\begin{thebibliography}{70}


\ifx \showCODEN    \undefined \def \showCODEN     #1{\unskip}     \fi
\ifx \showDOI      \undefined \def \showDOI       #1{#1}\fi
\ifx \showISBNx    \undefined \def \showISBNx     #1{\unskip}     \fi
\ifx \showISBNxiii \undefined \def \showISBNxiii  #1{\unskip}     \fi
\ifx \showISSN     \undefined \def \showISSN      #1{\unskip}     \fi
\ifx \showLCCN     \undefined \def \showLCCN      #1{\unskip}     \fi
\ifx \shownote     \undefined \def \shownote      #1{#1}          \fi
\ifx \showarticletitle \undefined \def \showarticletitle #1{#1}   \fi
\ifx \showURL      \undefined \def \showURL       {\relax}        \fi
\providecommand\bibfield[2]{#2}
\providecommand\bibinfo[2]{#2}
\providecommand\natexlab[1]{#1}
\providecommand\showeprint[2][]{arXiv:#2}

\bibitem[\protect\citeauthoryear{Anil, Pereyra, Passos, Ormandi, Dahl, and
  Hinton}{Anil et~al\mbox{.}}{2018}]%
        {anil2018large}
\bibfield{author}{\bibinfo{person}{Rohan Anil}, \bibinfo{person}{Gabriel
  Pereyra}, \bibinfo{person}{Alexandre Passos}, \bibinfo{person}{Robert
  Ormandi}, \bibinfo{person}{George~E Dahl}, {and} \bibinfo{person}{Geoffrey~E
  Hinton}.} \bibinfo{year}{2018}\natexlab{}.
\newblock \showarticletitle{Large scale distributed neural network training
  through online distillation}.
\newblock \bibinfo{journal}{\emph{ICLR}}.
\newblock


\bibitem[\protect\citeauthoryear{Bahri, Bahl, and Zafeiriou}{Bahri
  et~al\mbox{.}}{2021}]%
        {bahri2021binary}
\bibfield{author}{\bibinfo{person}{Mehdi Bahri}, \bibinfo{person}{Ga{\'e}tan
  Bahl}, {and} \bibinfo{person}{Stefanos Zafeiriou}.}
  \bibinfo{year}{2021}\natexlab{}.
\newblock \showarticletitle{Binary Graph Neural Networks}. In
  \bibinfo{booktitle}{\emph{CVPR}}. \bibinfo{pages}{9492--9501}.
\newblock


\bibitem[\protect\citeauthoryear{Banner, Hubara, Hoffer, and Soudry}{Banner
  et~al\mbox{.}}{2018}]%
        {banner2018scalable}
\bibfield{author}{\bibinfo{person}{Ron Banner}, \bibinfo{person}{Itay Hubara},
  \bibinfo{person}{Elad Hoffer}, {and} \bibinfo{person}{Daniel Soudry}.}
  \bibinfo{year}{2018}\natexlab{}.
\newblock \showarticletitle{Scalable methods for 8-bit training of neural
  networks}.
\newblock \bibinfo{journal}{\emph{NeurIPS}}  \bibinfo{volume}{31}.
\newblock


\bibitem[\protect\citeauthoryear{Bengio, L{\'e}onard, and Courville}{Bengio
  et~al\mbox{.}}{2013}]%
        {bengio2013estimating}
\bibfield{author}{\bibinfo{person}{Yoshua Bengio}, \bibinfo{person}{Nicholas
  L{\'e}onard}, {and} \bibinfo{person}{Aaron Courville}.}
  \bibinfo{year}{2013}\natexlab{}.
\newblock \showarticletitle{Estimating or propagating gradients through
  stochastic neurons for conditional computation}.
\newblock \bibinfo{journal}{\emph{arXiv}}.
\newblock


\bibitem[\protect\citeauthoryear{Berg, Kipf, and Welling}{Berg
  et~al\mbox{.}}{2017}]%
        {berg2017graph}
\bibfield{author}{\bibinfo{person}{Rianne van~den Berg},
  \bibinfo{person}{Thomas~N Kipf}, {and} \bibinfo{person}{Max Welling}.}
  \bibinfo{year}{2017}\natexlab{}.
\newblock \showarticletitle{Graph convolutional matrix completion}.
\newblock \bibinfo{journal}{\emph{arXiv preprint arXiv:1706.02263}}.
\newblock


\bibitem[\protect\citeauthoryear{Bracewell and Bracewell}{Bracewell and
  Bracewell}{1986}]%
        {bracewell1986fourier}
\bibfield{author}{\bibinfo{person}{Ronald~Newbold Bracewell} {and}
  \bibinfo{person}{Ronald~N Bracewell}.} \bibinfo{year}{1986}\natexlab{}.
\newblock \bibinfo{booktitle}{\emph{The Fourier transform and its
  applications}}. Vol.~\bibinfo{volume}{31999}.
\newblock \bibinfo{publisher}{McGraw-Hill New York}.
\newblock


\bibitem[\protect\citeauthoryear{Cao, Long, Wang, and Yu}{Cao
  et~al\mbox{.}}{2017}]%
        {hashnet}
\bibfield{author}{\bibinfo{person}{Zhangjie Cao}, \bibinfo{person}{Mingsheng
  Long}, \bibinfo{person}{Jianmin Wang}, {and} \bibinfo{person}{Philip~S Yu}.}
  \bibinfo{year}{2017}\natexlab{}.
\newblock \showarticletitle{Hashnet: Deep learning to hash by continuation}. In
  \bibinfo{booktitle}{\emph{ICCV}}. \bibinfo{pages}{5608--5617}.
\newblock


\bibitem[\protect\citeauthoryear{Chen, Zhang, He, Nie, Liu, and Chua}{Chen
  et~al\mbox{.}}{2017}]%
        {chen2017attentive}
\bibfield{author}{\bibinfo{person}{Jingyuan Chen}, \bibinfo{person}{Hanwang
  Zhang}, \bibinfo{person}{Xiangnan He}, \bibinfo{person}{Liqiang Nie},
  \bibinfo{person}{Wei Liu}, {and} \bibinfo{person}{Tat-Seng Chua}.}
  \bibinfo{year}{2017}\natexlab{}.
\newblock \showarticletitle{Attentive collaborative filtering: Multimedia
  recommendation with item-and component-level attention}. In
  \bibinfo{booktitle}{\emph{SIGIR}}. \bibinfo{pages}{335--344}.
\newblock


\bibitem[\protect\citeauthoryear{Chen, Yang, Zhang, Zhao, Meng, Hao, and
  King}{Chen et~al\mbox{.}}{2022b}]%
        {chen2021modeling}
\bibfield{author}{\bibinfo{person}{Yankai Chen}, \bibinfo{person}{Menglin
  Yang}, \bibinfo{person}{Yingxue Zhang}, \bibinfo{person}{Mengchen Zhao},
  \bibinfo{person}{Ziqiao Meng}, \bibinfo{person}{Jianye Hao}, {and}
  \bibinfo{person}{Irwin King}.} \bibinfo{year}{2022}\natexlab{b}.
\newblock \showarticletitle{Modeling Scale-free Graphs with Hyperbolic Geometry
  for Knowledge-aware Recommendation}.
\newblock \bibinfo{journal}{\emph{WSDM}}.
\newblock


\bibitem[\protect\citeauthoryear{Chen, Yang, Wang, Bai, Song, and King}{Chen
  et~al\mbox{.}}{2022a}]%
        {chen2021attentive}
\bibfield{author}{\bibinfo{person}{Yankai Chen}, \bibinfo{person}{Yaming Yang},
  \bibinfo{person}{Yujing Wang}, \bibinfo{person}{Jing Bai},
  \bibinfo{person}{Xiangchen Song}, {and} \bibinfo{person}{Irwin King}.}
  \bibinfo{year}{2022}\natexlab{a}.
\newblock \showarticletitle{Attentive Knowledge-aware Graph Convolutional
  Networks with Collaborative Guidance for Personalized Recommendation}.
\newblock \bibinfo{journal}{\emph{ICDE}}.
\newblock


\bibitem[\protect\citeauthoryear{Covington, Adams, and Sargin}{Covington
  et~al\mbox{.}}{2016}]%
        {covington2016deep}
\bibfield{author}{\bibinfo{person}{Paul Covington}, \bibinfo{person}{Jay
  Adams}, {and} \bibinfo{person}{Emre Sargin}.}
  \bibinfo{year}{2016}\natexlab{}.
\newblock \showarticletitle{Deep neural networks for youtube recommendations}.
  In \bibinfo{booktitle}{\emph{Recsys}}. \bibinfo{pages}{191--198}.
\newblock


\bibitem[\protect\citeauthoryear{Darabi, Belbahri, Courbariaux, and Nia}{Darabi
  et~al\mbox{.}}{2018}]%
        {darabi2018bnn}
\bibfield{author}{\bibinfo{person}{Sajad Darabi}, \bibinfo{person}{Mouloud
  Belbahri}, \bibinfo{person}{Matthieu Courbariaux}, {and}
  \bibinfo{person}{Vahid~Partovi Nia}.} \bibinfo{year}{2018}\natexlab{}.
\newblock \showarticletitle{Bnn+: Improved binary network training}.
\newblock \bibinfo{journal}{\emph{arXiv}}.
\newblock


\bibitem[\protect\citeauthoryear{Erin~Liong, Lu, Wang, Moulin, and
  Zhou}{Erin~Liong et~al\mbox{.}}{2015}]%
        {erin2015deep}
\bibfield{author}{\bibinfo{person}{Venice Erin~Liong}, \bibinfo{person}{Jiwen
  Lu}, \bibinfo{person}{Gang Wang}, \bibinfo{person}{Pierre Moulin}, {and}
  \bibinfo{person}{Jie Zhou}.} \bibinfo{year}{2015}\natexlab{}.
\newblock \showarticletitle{Deep hashing for compact binary codes learning}. In
  \bibinfo{booktitle}{\emph{CVPR}}. \bibinfo{pages}{2475--2483}.
\newblock


\bibitem[\protect\citeauthoryear{function}{function}{2022}]%
        {step_function}
\bibfield{author}{\bibinfo{person}{Step function}.}
  \bibinfo{year}{2022}\natexlab{}.
\newblock
\newblock
\newblock
\shownote{\url{https://en.wikipedia.org/wiki/Heaviside_step_function}.}


\bibitem[\protect\citeauthoryear{Geng, Zhang, Bian, and Chua}{Geng
  et~al\mbox{.}}{2015}]%
        {geng2015learning}
\bibfield{author}{\bibinfo{person}{Xue Geng}, \bibinfo{person}{Hanwang Zhang},
  \bibinfo{person}{Jingwen Bian}, {and} \bibinfo{person}{Tat-Seng Chua}.}
  \bibinfo{year}{2015}\natexlab{}.
\newblock \showarticletitle{Learning image and user features for recommendation
  in social networks}. In \bibinfo{booktitle}{\emph{ICCV}}.
\newblock


\bibitem[\protect\citeauthoryear{Gionis, Indyk, Motwani, et~al\mbox{.}}{Gionis
  et~al\mbox{.}}{1999}]%
        {lsh}
\bibfield{author}{\bibinfo{person}{Aristides Gionis}, \bibinfo{person}{Piotr
  Indyk}, \bibinfo{person}{Rajeev Motwani}, {et~al\mbox{.}}}
  \bibinfo{year}{1999}\natexlab{}.
\newblock \showarticletitle{Similarity search in high dimensions via hashing}.
  In \bibinfo{booktitle}{\emph{VLDB}}, Vol.~\bibinfo{volume}{99}.
  \bibinfo{pages}{518--529}.
\newblock


\bibitem[\protect\citeauthoryear{Gong, Liu, Jiang, Li, Hu, Lin, Yu, and
  Yan}{Gong et~al\mbox{.}}{2019}]%
        {gong2019differentiable}
\bibfield{author}{\bibinfo{person}{Ruihao Gong}, \bibinfo{person}{Xianglong
  Liu}, \bibinfo{person}{Shenghu Jiang}, \bibinfo{person}{Tianxiang Li},
  \bibinfo{person}{Peng Hu}, \bibinfo{person}{Jiazhen Lin},
  \bibinfo{person}{Fengwei Yu}, {and} \bibinfo{person}{Junjie Yan}.}
  \bibinfo{year}{2019}\natexlab{}.
\newblock \showarticletitle{Differentiable soft quantization: Bridging
  full-precision and low-bit neural networks}. In
  \bibinfo{booktitle}{\emph{ICCV}}. \bibinfo{pages}{4852--4861}.
\newblock


\bibitem[\protect\citeauthoryear{Hamilton, Ying, and Leskovec}{Hamilton
  et~al\mbox{.}}{2017}]%
        {graphsage}
\bibfield{author}{\bibinfo{person}{William~L Hamilton}, \bibinfo{person}{Rex
  Ying}, {and} \bibinfo{person}{Jure Leskovec}.}
  \bibinfo{year}{2017}\natexlab{}.
\newblock \showarticletitle{Inductive representation learning on large graphs}.
  In \bibinfo{booktitle}{\emph{NeurIPS}}. \bibinfo{pages}{1025--1035}.
\newblock


\bibitem[\protect\citeauthoryear{H{\aa}stad}{H{\aa}stad}{2001}]%
        {haastad2001some}
\bibfield{author}{\bibinfo{person}{Johan H{\aa}stad}.}
  \bibinfo{year}{2001}\natexlab{}.
\newblock \showarticletitle{Some optimal inapproximability results}.
\newblock \bibinfo{journal}{\emph{Journal of the ACM (JACM)}}
  \bibinfo{volume}{48}, \bibinfo{number}{4}, \bibinfo{pages}{798--859}.
\newblock


\bibitem[\protect\citeauthoryear{He and McAuley}{He and McAuley}{2016}]%
        {he2016ups}
\bibfield{author}{\bibinfo{person}{Ruining He} {and} \bibinfo{person}{Julian
  McAuley}.} \bibinfo{year}{2016}\natexlab{}.
\newblock \showarticletitle{Modeling the visual evolution of fashion trends
  with one-class collaborative filtering}. In \bibinfo{booktitle}{\emph{WWW}}.
  \bibinfo{pages}{507--517}.
\newblock


\bibitem[\protect\citeauthoryear{He, Deng, Wang, Li, Zhang, and Wang}{He
  et~al\mbox{.}}{2020}]%
        {lightgcn}
\bibfield{author}{\bibinfo{person}{Xiangnan He}, \bibinfo{person}{Kuan Deng},
  \bibinfo{person}{Xiang Wang}, \bibinfo{person}{Yan Li},
  \bibinfo{person}{Yongdong Zhang}, {and} \bibinfo{person}{Meng Wang}.}
  \bibinfo{year}{2020}\natexlab{}.
\newblock \showarticletitle{Lightgcn: Simplifying and powering graph
  convolution network for recommendation}. In
  \bibinfo{booktitle}{\emph{SIGIR}}. \bibinfo{pages}{639--648}.
\newblock


\bibitem[\protect\citeauthoryear{He, He, Song, Liu, Jiang, and Chua}{He
  et~al\mbox{.}}{2018}]%
        {he2018nais}
\bibfield{author}{\bibinfo{person}{Xiangnan He}, \bibinfo{person}{Zhankui He},
  \bibinfo{person}{Jingkuan Song}, \bibinfo{person}{Zhenguang Liu},
  \bibinfo{person}{Yu-Gang Jiang}, {and} \bibinfo{person}{Tat-Seng Chua}.}
  \bibinfo{year}{2018}\natexlab{}.
\newblock \showarticletitle{Nais: Neural attentive item similarity model for
  recommendation}.
\newblock \bibinfo{journal}{\emph{TKDE}} \bibinfo{volume}{30},
  \bibinfo{number}{12}, \bibinfo{pages}{2354--2366}.
\newblock


\bibitem[\protect\citeauthoryear{He, Liao, Zhang, Nie, Hu, and Chua}{He
  et~al\mbox{.}}{2017}]%
        {neurcf}
\bibfield{author}{\bibinfo{person}{Xiangnan He}, \bibinfo{person}{Lizi Liao},
  \bibinfo{person}{Hanwang Zhang}, \bibinfo{person}{Liqiang Nie},
  \bibinfo{person}{Xia Hu}, {and} \bibinfo{person}{Tat-Seng Chua}.}
  \bibinfo{year}{2017}\natexlab{}.
\newblock \showarticletitle{Neural collaborative filtering}. In
  \bibinfo{booktitle}{\emph{WWW}}. \bibinfo{pages}{173--182}.
\newblock


\bibitem[\protect\citeauthoryear{He, Zhang, Kan, and Chua}{He
  et~al\mbox{.}}{2016}]%
        {he2016fast}
\bibfield{author}{\bibinfo{person}{Xiangnan He}, \bibinfo{person}{Hanwang
  Zhang}, \bibinfo{person}{Min-Yen Kan}, {and} \bibinfo{person}{Tat-Seng
  Chua}.} \bibinfo{year}{2016}\natexlab{}.
\newblock \showarticletitle{Fast matrix factorization for recommendation with
  implicit feedback}. In \bibinfo{booktitle}{\emph{SIGIR}}.
\newblock


\bibitem[\protect\citeauthoryear{Hinton, Vinyals, and Dean}{Hinton
  et~al\mbox{.}}{2015}]%
        {hinton2015distilling}
\bibfield{author}{\bibinfo{person}{Geoffrey Hinton}, \bibinfo{person}{Oriol
  Vinyals}, {and} \bibinfo{person}{Jeff Dean}.}
  \bibinfo{year}{2015}\natexlab{}.
\newblock \showarticletitle{Distilling the knowledge in a neural network}.
\newblock \bibinfo{journal}{\emph{arXiv preprint arXiv:1503.02531}}.
\newblock


\bibitem[\protect\citeauthoryear{Jang, Gu, and Poole}{Jang
  et~al\mbox{.}}{2017}]%
        {gumbel1}
\bibfield{author}{\bibinfo{person}{Eric Jang}, \bibinfo{person}{Shixiang Gu},
  {and} \bibinfo{person}{Ben Poole}.} \bibinfo{year}{2017}\natexlab{}.
\newblock \showarticletitle{Categorical reparameterization with
  gumbel-softmax}. In \bibinfo{booktitle}{\emph{5th ICLR}}.
\newblock


\bibitem[\protect\citeauthoryear{Kang, Cheng, Yao, Yi, Chen, Hong, and
  Chi}{Kang et~al\mbox{.}}{2021}]%
        {kang2021learning}
\bibfield{author}{\bibinfo{person}{Wang-Cheng Kang},
  \bibinfo{person}{Derek~Zhiyuan Cheng}, \bibinfo{person}{Tiansheng Yao},
  \bibinfo{person}{Xinyang Yi}, \bibinfo{person}{Ting Chen},
  \bibinfo{person}{Lichan Hong}, {and} \bibinfo{person}{Ed~H Chi}.}
  \bibinfo{year}{2021}\natexlab{}.
\newblock \showarticletitle{Learning to embed categorical features without
  embedding tables for recommendation}.
\newblock \bibinfo{journal}{\emph{SIGKDD}}.
\newblock


\bibitem[\protect\citeauthoryear{Kang and McAuley}{Kang and McAuley}{2019}]%
        {kang2019candidate}
\bibfield{author}{\bibinfo{person}{Wang-Cheng Kang} {and}
  \bibinfo{person}{Julian McAuley}.} \bibinfo{year}{2019}\natexlab{}.
\newblock \showarticletitle{Candidate generation with binary codes for
  large-scale top-n recommendation}. In \bibinfo{booktitle}{\emph{CIKM}}.
  \bibinfo{pages}{1523--1532}.
\newblock


\bibitem[\protect\citeauthoryear{Kingma and Ba}{Kingma and Ba}{2015}]%
        {adam}
\bibfield{author}{\bibinfo{person}{Diederik~P Kingma} {and}
  \bibinfo{person}{Jimmy Ba}.} \bibinfo{year}{2015}\natexlab{}.
\newblock \showarticletitle{Adam: A method for stochastic optimization}. In
  \bibinfo{booktitle}{\emph{ICLR}}.
\newblock


\bibitem[\protect\citeauthoryear{Kipf and Welling}{Kipf and Welling}{2017}]%
        {kipf2016semi}
\bibfield{author}{\bibinfo{person}{Thomas~N Kipf} {and} \bibinfo{person}{Max
  Welling}.} \bibinfo{year}{2017}\natexlab{}.
\newblock \showarticletitle{Semi-supervised classification with graph
  convolutional networks}. In \bibinfo{booktitle}{\emph{5th ICLR}}.
\newblock


\bibitem[\protect\citeauthoryear{Koh and Liang}{Koh and Liang}{2017}]%
        {koh2017understanding}
\bibfield{author}{\bibinfo{person}{Pang~Wei Koh} {and} \bibinfo{person}{Percy
  Liang}.} \bibinfo{year}{2017}\natexlab{}.
\newblock \showarticletitle{Understanding black-box predictions via influence
  functions}. In \bibinfo{booktitle}{\emph{ICML}}. PMLR,
  \bibinfo{pages}{1885--1894}.
\newblock


\bibitem[\protect\citeauthoryear{Koren, Bell, and Volinsky}{Koren
  et~al\mbox{.}}{2009}]%
        {koren2009matrix}
\bibfield{author}{\bibinfo{person}{Yehuda Koren}, \bibinfo{person}{Robert
  Bell}, {and} \bibinfo{person}{Chris Volinsky}.}
  \bibinfo{year}{2009}\natexlab{}.
\newblock \showarticletitle{Matrix factorization techniques for recommender
  systems}.
\newblock \bibinfo{journal}{\emph{Computer}} \bibinfo{volume}{42},
  \bibinfo{number}{8}, \bibinfo{pages}{30--37}.
\newblock


\bibitem[\protect\citeauthoryear{Li, Muller, Thabet, and Ghanem}{Li
  et~al\mbox{.}}{2019}]%
        {li2019deepgcns}
\bibfield{author}{\bibinfo{person}{Guohao Li}, \bibinfo{person}{Matthias
  Muller}, \bibinfo{person}{Ali Thabet}, {and} \bibinfo{person}{Bernard
  Ghanem}.} \bibinfo{year}{2019}\natexlab{}.
\newblock \showarticletitle{Deepgcns: Can gcns go as deep as cnns?}. In
  \bibinfo{booktitle}{\emph{ICCV}}. \bibinfo{pages}{9267--9276}.
\newblock


\bibitem[\protect\citeauthoryear{Li, Liu, and Ji}{Li et~al\mbox{.}}{2014}]%
        {li2014two}
\bibfield{author}{\bibinfo{person}{Hao Li}, \bibinfo{person}{Wei Liu}, {and}
  \bibinfo{person}{Heng Ji}.} \bibinfo{year}{2014}\natexlab{}.
\newblock \showarticletitle{Two-Stage Hashing for Fast Document Retrieval.}. In
  \bibinfo{booktitle}{\emph{ACL (2)}}. \bibinfo{pages}{495--500}.
\newblock


\bibitem[\protect\citeauthoryear{Li, Han, and Wu}{Li et~al\mbox{.}}{2018}]%
        {li2018deeper}
\bibfield{author}{\bibinfo{person}{Qimai Li}, \bibinfo{person}{Zhichao Han},
  {and} \bibinfo{person}{Xiao-Ming Wu}.} \bibinfo{year}{2018}\natexlab{}.
\newblock \showarticletitle{Deeper insights into graph convolutional networks
  for semi-supervised learning}. In \bibinfo{booktitle}{\emph{AAAI}}.
\newblock


\bibitem[\protect\citeauthoryear{Liang, Charlin, McInerney, and Blei}{Liang
  et~al\mbox{.}}{2016}]%
        {liang2016modeling}
\bibfield{author}{\bibinfo{person}{Dawen Liang}, \bibinfo{person}{Laurent
  Charlin}, \bibinfo{person}{James McInerney}, {and} \bibinfo{person}{David~M
  Blei}.} \bibinfo{year}{2016}\natexlab{}.
\newblock \showarticletitle{Modeling user exposure in recommendation}. In
  \bibinfo{booktitle}{\emph{WWW}}. \bibinfo{pages}{951--961}.
\newblock


\bibitem[\protect\citeauthoryear{Lin, Zhao, and Pan}{Lin et~al\mbox{.}}{2017}]%
        {lin2017towards}
\bibfield{author}{\bibinfo{person}{Xiaofan Lin}, \bibinfo{person}{Cong Zhao},
  {and} \bibinfo{person}{Wei Pan}.} \bibinfo{year}{2017}\natexlab{}.
\newblock \showarticletitle{Towards accurate binary convolutional neural
  network}. In \bibinfo{booktitle}{\emph{NeurIPS}}.
\newblock


\bibitem[\protect\citeauthoryear{Liu, Ding, Xia, Hu, Zhang, Liu, Zhuang, and
  Guo}{Liu et~al\mbox{.}}{2019}]%
        {RBCN}
\bibfield{author}{\bibinfo{person}{Chunlei Liu}, \bibinfo{person}{Wenrui Ding},
  \bibinfo{person}{Xin Xia}, \bibinfo{person}{Yuan Hu},
  \bibinfo{person}{Baochang Zhang}, \bibinfo{person}{Jianzhuang Liu},
  \bibinfo{person}{Bohan Zhuang}, {and} \bibinfo{person}{Guodong Guo}.}
  \bibinfo{year}{2019}\natexlab{}.
\newblock \showarticletitle{RBCN: Rectified binary convolutional networks for
  enhancing the performance of 1-bit DCNNs}.
\newblock \bibinfo{journal}{\emph{arXiv}}.
\newblock


\bibitem[\protect\citeauthoryear{Liu, He, Deng, and Lang}{Liu
  et~al\mbox{.}}{2014}]%
        {liu2014collaborative}
\bibfield{author}{\bibinfo{person}{Xianglong Liu}, \bibinfo{person}{Junfeng
  He}, \bibinfo{person}{Cheng Deng}, {and} \bibinfo{person}{Bo Lang}.}
  \bibinfo{year}{2014}\natexlab{}.
\newblock \showarticletitle{Collaborative hashing}. In
  \bibinfo{booktitle}{\emph{CVPR}}. \bibinfo{pages}{2139--2146}.
\newblock


\bibitem[\protect\citeauthoryear{Maddison, Mnih, and Teh}{Maddison
  et~al\mbox{.}}{2017}]%
        {gumbel2}
\bibfield{author}{\bibinfo{person}{Chris~J Maddison}, \bibinfo{person}{Andriy
  Mnih}, {and} \bibinfo{person}{Yee~Whye Teh}.}
  \bibinfo{year}{2017}\natexlab{}.
\newblock \showarticletitle{The concrete distribution: A continuous relaxation
  of discrete random variables}. In \bibinfo{booktitle}{\emph{5th ICLR}}.
\newblock


\bibitem[\protect\citeauthoryear{Park and Tuzhilin}{Park and Tuzhilin}{2008}]%
        {park2008long}
\bibfield{author}{\bibinfo{person}{Yoon-Joo Park} {and}
  \bibinfo{person}{Alexander Tuzhilin}.} \bibinfo{year}{2008}\natexlab{}.
\newblock \showarticletitle{The long tail of recommender systems and how to
  leverage it}. In \bibinfo{booktitle}{\emph{RecSys}}. \bibinfo{pages}{11--18}.
\newblock


\bibitem[\protect\citeauthoryear{product statistics}{product
  statistics}{2022}]%
        {amazon_2}
\bibfield{author}{\bibinfo{person}{Amazon product statistics}.}
  \bibinfo{year}{2022}\natexlab{}.
\newblock
\newblock
\newblock
\shownote{\url{https://www.retailtouchpoints.com/resources/how-many-products-does-amazon-carry}.}


\bibitem[\protect\citeauthoryear{Qin, Gong, Liu, Shen, Wei, Yu, and Song}{Qin
  et~al\mbox{.}}{2020}]%
        {qin2020forward}
\bibfield{author}{\bibinfo{person}{Haotong Qin}, \bibinfo{person}{Ruihao Gong},
  \bibinfo{person}{Xianglong Liu}, \bibinfo{person}{Mingzhu Shen},
  \bibinfo{person}{Ziran Wei}, \bibinfo{person}{Fengwei Yu}, {and}
  \bibinfo{person}{Jingkuan Song}.} \bibinfo{year}{2020}\natexlab{}.
\newblock \showarticletitle{Forward and backward information retention for
  accurate binary neural networks}. In \bibinfo{booktitle}{\emph{CVPR}}.
  \bibinfo{pages}{2250--2259}.
\newblock


\bibitem[\protect\citeauthoryear{Rastegari, Ordonez, Redmon, and
  Farhadi}{Rastegari et~al\mbox{.}}{2016}]%
        {rastegari2016xnor}
\bibfield{author}{\bibinfo{person}{Mohammad Rastegari},
  \bibinfo{person}{Vicente Ordonez}, \bibinfo{person}{Joseph Redmon}, {and}
  \bibinfo{person}{Ali Farhadi}.} \bibinfo{year}{2016}\natexlab{}.
\newblock \showarticletitle{Xnor-net: Imagenet classification using binary
  convolutional neural networks}. In \bibinfo{booktitle}{\emph{ECCV}}.
  Springer, \bibinfo{pages}{525--542}.
\newblock


\bibitem[\protect\citeauthoryear{Rendle and Freudenthaler}{Rendle and
  Freudenthaler}{2014}]%
        {rendle2014improving}
\bibfield{author}{\bibinfo{person}{Steffen Rendle} {and}
  \bibinfo{person}{Christoph Freudenthaler}.} \bibinfo{year}{2014}\natexlab{}.
\newblock \showarticletitle{Improving pairwise learning for item recommendation
  from implicit feedback}. In \bibinfo{booktitle}{\emph{WSDM}}.
  \bibinfo{pages}{273--282}.
\newblock


\bibitem[\protect\citeauthoryear{Rendle, Freudenthaler, Gantner, and
  Schmidt-Thieme}{Rendle et~al\mbox{.}}{2012}]%
        {rendle2012bpr}
\bibfield{author}{\bibinfo{person}{Steffen Rendle}, \bibinfo{person}{Christoph
  Freudenthaler}, \bibinfo{person}{Zeno Gantner}, {and} \bibinfo{person}{Lars
  Schmidt-Thieme}.} \bibinfo{year}{2012}\natexlab{}.
\newblock \showarticletitle{BPR: Bayesian personalized ranking from implicit
  feedback}.
\newblock \bibinfo{journal}{\emph{arXiv}}.
\newblock


\bibitem[\protect\citeauthoryear{Song, Meng, Zhang, and King}{Song
  et~al\mbox{.}}{2021}]%
        {song2021semi}
\bibfield{author}{\bibinfo{person}{Zixing Song}, \bibinfo{person}{Ziqiao Meng},
  \bibinfo{person}{Yifei Zhang}, {and} \bibinfo{person}{Irwin King}.}
  \bibinfo{year}{2021}\natexlab{}.
\newblock \showarticletitle{Semi-supervised Multi-label Learning for
  Graph-structured Data}. In \bibinfo{booktitle}{\emph{CIKM}}.
  \bibinfo{publisher}{{ACM}}, \bibinfo{pages}{1723--1733}.
\newblock


\bibitem[\protect\citeauthoryear{Song, Yang, Xu, and King}{Song
  et~al\mbox{.}}{2022}]%
        {song2022graph}
\bibfield{author}{\bibinfo{person}{Zixing Song}, \bibinfo{person}{Xiangli
  Yang}, \bibinfo{person}{Zenglin Xu}, {and} \bibinfo{person}{Irwin King}.}
  \bibinfo{year}{2022}\natexlab{}.
\newblock \showarticletitle{Graph-based semi-supervised learning: A
  comprehensive review}.
\newblock \bibinfo{journal}{\emph{TNNLS}}.
\newblock


\bibitem[\protect\citeauthoryear{Tailor, Fernandez-Marques, and Lane}{Tailor
  et~al\mbox{.}}{2021}]%
        {tailor2020degree}
\bibfield{author}{\bibinfo{person}{Shyam~A Tailor}, \bibinfo{person}{Javier
  Fernandez-Marques}, {and} \bibinfo{person}{Nicholas~D Lane}.}
  \bibinfo{year}{2021}\natexlab{}.
\newblock \showarticletitle{Degree-quant: Quantization-aware training for graph
  neural networks}.
\newblock \bibinfo{journal}{\emph{9th ICLR}}.
\newblock


\bibitem[\protect\citeauthoryear{Tan, Liu, Zhao, Yang, Zhou, and Hu}{Tan
  et~al\mbox{.}}{2020}]%
        {hashgnn}
\bibfield{author}{\bibinfo{person}{Qiaoyu Tan}, \bibinfo{person}{Ninghao Liu},
  \bibinfo{person}{Xing Zhao}, \bibinfo{person}{Hongxia Yang},
  \bibinfo{person}{Jingren Zhou}, {and} \bibinfo{person}{Xia Hu}.}
  \bibinfo{year}{2020}\natexlab{}.
\newblock \showarticletitle{Learning to hash with GNNs for recommender
  systems}. In \bibinfo{booktitle}{\emph{WWW}}. \bibinfo{pages}{1988--1998}.
\newblock


\bibitem[\protect\citeauthoryear{Tang and Wang}{Tang and Wang}{2018}]%
        {tang2018ranking}
\bibfield{author}{\bibinfo{person}{Jiaxi Tang} {and} \bibinfo{person}{Ke
  Wang}.} \bibinfo{year}{2018}\natexlab{}.
\newblock \showarticletitle{Learning compact ranking models with high
  performance for recommender system}. In \bibinfo{booktitle}{\emph{SIGKDD}}.
  \bibinfo{pages}{2289--2298}.
\newblock


\bibitem[\protect\citeauthoryear{user statistics}{user statistics}{2022}]%
        {amazon}
\bibfield{author}{\bibinfo{person}{Amazon user statistics}.}
  \bibinfo{year}{2022}\natexlab{}.
\newblock
\newblock
\newblock
\shownote{\url{https://backlinko.com/amazon-prime-users}.}


\bibitem[\protect\citeauthoryear{Veli{\v{c}}kovi{\'c}, Cucurull, Casanova,
  Romero, Lio, and Bengio}{Veli{\v{c}}kovi{\'c} et~al\mbox{.}}{2018}]%
        {velivckovic2017graph}
\bibfield{author}{\bibinfo{person}{Petar Veli{\v{c}}kovi{\'c}},
  \bibinfo{person}{Guillem Cucurull}, \bibinfo{person}{Arantxa Casanova},
  \bibinfo{person}{Adriana Romero}, \bibinfo{person}{Pietro Lio}, {and}
  \bibinfo{person}{Yoshua Bengio}.} \bibinfo{year}{2018}\natexlab{}.
\newblock \showarticletitle{Graph attention networks}.
\newblock \bibinfo{journal}{\emph{ICLR}}.
\newblock


\bibitem[\protect\citeauthoryear{Wang, Wang, Yang, Yang, and Guo}{Wang
  et~al\mbox{.}}{2021}]%
        {bigcn}
\bibfield{author}{\bibinfo{person}{Junfu Wang}, \bibinfo{person}{Yunhong Wang},
  \bibinfo{person}{Zhen Yang}, \bibinfo{person}{Liang Yang}, {and}
  \bibinfo{person}{Yuanfang Guo}.} \bibinfo{year}{2021}\natexlab{}.
\newblock \showarticletitle{Bi-gcn: Binary graph convolutional network}. In
  \bibinfo{booktitle}{\emph{CVPR}}. \bibinfo{pages}{1561--1570}.
\newblock


\bibitem[\protect\citeauthoryear{Wang, Zhang, Sebe, Shen, et~al\mbox{.}}{Wang
  et~al\mbox{.}}{2017}]%
        {wang2017survey}
\bibfield{author}{\bibinfo{person}{Jingdong Wang}, \bibinfo{person}{Ting
  Zhang}, \bibinfo{person}{Nicu Sebe}, \bibinfo{person}{Heng~Tao Shen},
  {et~al\mbox{.}}} \bibinfo{year}{2017}\natexlab{}.
\newblock \showarticletitle{A survey on learning to hash}.
\newblock \bibinfo{journal}{\emph{TPAMI}} \bibinfo{volume}{40},
  \bibinfo{number}{4}, \bibinfo{pages}{769--790}.
\newblock


\bibitem[\protect\citeauthoryear{Wang, He, Wang, Feng, and Chua}{Wang
  et~al\mbox{.}}{2019}]%
        {ngcf}
\bibfield{author}{\bibinfo{person}{Xiang Wang}, \bibinfo{person}{Xiangnan He},
  \bibinfo{person}{Meng Wang}, \bibinfo{person}{Fuli Feng}, {and}
  \bibinfo{person}{Tat-Seng Chua}.} \bibinfo{year}{2019}\natexlab{}.
\newblock \showarticletitle{Neural graph collaborative filtering}. In
  \bibinfo{booktitle}{\emph{SIGIR}}. \bibinfo{pages}{165--174}.
\newblock


\bibitem[\protect\citeauthoryear{Wang, Jin, Zhang, He, Xu, and Chua}{Wang
  et~al\mbox{.}}{2020}]%
        {dgcf}
\bibfield{author}{\bibinfo{person}{Xiang Wang}, \bibinfo{person}{Hongye Jin},
  \bibinfo{person}{An Zhang}, \bibinfo{person}{Xiangnan He},
  \bibinfo{person}{Tong Xu}, {and} \bibinfo{person}{Tat-Seng Chua}.}
  \bibinfo{year}{2020}\natexlab{}.
\newblock \showarticletitle{Disentangled graph collaborative filtering}. In
  \bibinfo{booktitle}{\emph{SIGIR}}. \bibinfo{pages}{1001--1010}.
\newblock


\bibitem[\protect\citeauthoryear{Wu, Souza, Zhang, Fifty, Yu, and
  Weinberger}{Wu et~al\mbox{.}}{2019}]%
        {wu2019simplifying}
\bibfield{author}{\bibinfo{person}{Felix Wu}, \bibinfo{person}{Amauri Souza},
  \bibinfo{person}{Tianyi Zhang}, \bibinfo{person}{Christopher Fifty},
  \bibinfo{person}{Tao Yu}, {and} \bibinfo{person}{Kilian Weinberger}.}
  \bibinfo{year}{2019}\natexlab{}.
\newblock \showarticletitle{Simplifying graph convolutional networks}. In
  \bibinfo{booktitle}{\emph{ICML}}. PMLR.
\newblock


\bibitem[\protect\citeauthoryear{Wu, Pan, Chen, Long, Zhang, and Philip}{Wu
  et~al\mbox{.}}{2020}]%
        {wu2020comprehensive}
\bibfield{author}{\bibinfo{person}{Zonghan Wu}, \bibinfo{person}{Shirui Pan},
  \bibinfo{person}{Fengwen Chen}, \bibinfo{person}{Guodong Long},
  \bibinfo{person}{Chengqi Zhang}, {and} \bibinfo{person}{S~Yu Philip}.}
  \bibinfo{year}{2020}\natexlab{}.
\newblock \showarticletitle{A comprehensive survey on graph neural networks}.
\newblock \bibinfo{journal}{\emph{IEEE TNNLS}} \bibinfo{volume}{32},
  \bibinfo{number}{1}, \bibinfo{pages}{4--24}.
\newblock


\bibitem[\protect\citeauthoryear{Xie, Luong, Hovy, and Le}{Xie
  et~al\mbox{.}}{2020}]%
        {xie2020self}
\bibfield{author}{\bibinfo{person}{Qizhe Xie}, \bibinfo{person}{Minh-Thang
  Luong}, \bibinfo{person}{Eduard Hovy}, {and} \bibinfo{person}{Quoc~V Le}.}
  \bibinfo{year}{2020}\natexlab{}.
\newblock \showarticletitle{Self-training with noisy student improves imagenet
  classification}. In \bibinfo{booktitle}{\emph{CVPR}}.
  \bibinfo{pages}{10687--10698}.
\newblock


\bibitem[\protect\citeauthoryear{Xu, Li, Tian, Sonobe, Kawarabayashi, and
  Jegelka}{Xu et~al\mbox{.}}{2018}]%
        {xu2018representation}
\bibfield{author}{\bibinfo{person}{Keyulu Xu}, \bibinfo{person}{Chengtao Li},
  \bibinfo{person}{Yonglong Tian}, \bibinfo{person}{Tomohiro Sonobe},
  \bibinfo{person}{Ken-ichi Kawarabayashi}, {and} \bibinfo{person}{Stefanie
  Jegelka}.} \bibinfo{year}{2018}\natexlab{}.
\newblock \showarticletitle{Representation learning on graphs with jumping
  knowledge networks}. In \bibinfo{booktitle}{\emph{ICML}}. PMLR,
  \bibinfo{pages}{5453--5462}.
\newblock


\bibitem[\protect\citeauthoryear{Yang, Shen, Xing, Tian, Li, Deng, Huang, and
  Hua}{Yang et~al\mbox{.}}{2019}]%
        {sigmoid}
\bibfield{author}{\bibinfo{person}{Jiwei Yang}, \bibinfo{person}{Xu Shen},
  \bibinfo{person}{Jun Xing}, \bibinfo{person}{Xinmei Tian},
  \bibinfo{person}{Houqiang Li}, \bibinfo{person}{Bing Deng},
  \bibinfo{person}{Jianqiang Huang}, {and} \bibinfo{person}{Xian-sheng Hua}.}
  \bibinfo{year}{2019}\natexlab{}.
\newblock \showarticletitle{Quantization networks}. In
  \bibinfo{booktitle}{\emph{CVPR}}. \bibinfo{pages}{7308--7316}.
\newblock


\bibitem[\protect\citeauthoryear{Yang, Zhou, Kalander, Huang, and King}{Yang
  et~al\mbox{.}}{2021}]%
        {yang2021discrete}
\bibfield{author}{\bibinfo{person}{Menglin Yang}, \bibinfo{person}{Min Zhou},
  \bibinfo{person}{Marcus Kalander}, \bibinfo{person}{Zengfeng Huang}, {and}
  \bibinfo{person}{Irwin King}.} \bibinfo{year}{2021}\natexlab{}.
\newblock \showarticletitle{Discrete-time Temporal Network Embedding via
  Implicit Hierarchical Learning in Hyperbolic Space}. In
  \bibinfo{booktitle}{\emph{SIGKDD}}. \bibinfo{pages}{1975--1985}.
\newblock


\bibitem[\protect\citeauthoryear{Yang, Zhou, Liu, Lian, and King}{Yang
  et~al\mbox{.}}{2022}]%
        {yang2022hrcf}
\bibfield{author}{\bibinfo{person}{Menglin Yang}, \bibinfo{person}{Min Zhou},
  \bibinfo{person}{Jiahong Liu}, \bibinfo{person}{Defu Lian}, {and}
  \bibinfo{person}{Irwin King}.} \bibinfo{year}{2022}\natexlab{}.
\newblock \showarticletitle{HRCF: Enhancing collaborative filtering via
  hyperbolic geometric regularization}. In \bibinfo{booktitle}{\emph{WebConf}}.
  \bibinfo{pages}{2462--2471}.
\newblock


\bibitem[\protect\citeauthoryear{Ying, He, Chen, Eksombatchai, Hamilton, and
  Leskovec}{Ying et~al\mbox{.}}{2018}]%
        {pinsage}
\bibfield{author}{\bibinfo{person}{Rex Ying}, \bibinfo{person}{Ruining He},
  \bibinfo{person}{Kaifeng Chen}, \bibinfo{person}{Pong Eksombatchai},
  \bibinfo{person}{William~L Hamilton}, {and} \bibinfo{person}{Jure Leskovec}.}
  \bibinfo{year}{2018}\natexlab{}.
\newblock \showarticletitle{Graph convolutional neural networks for web-scale
  recommender systems}. In \bibinfo{booktitle}{\emph{SIGKDD}}.
  \bibinfo{pages}{974--983}.
\newblock


\bibitem[\protect\citeauthoryear{Zhang, Shen, Liu, He, Luan, and Chua}{Zhang
  et~al\mbox{.}}{2016}]%
        {zhang2016discrete}
\bibfield{author}{\bibinfo{person}{Hanwang Zhang}, \bibinfo{person}{Fumin
  Shen}, \bibinfo{person}{Wei Liu}, \bibinfo{person}{Xiangnan He},
  \bibinfo{person}{Huanbo Luan}, {and} \bibinfo{person}{Tat-Seng Chua}.}
  \bibinfo{year}{2016}\natexlab{}.
\newblock \showarticletitle{Discrete collaborative filtering}. In
  \bibinfo{booktitle}{\emph{SIGIR}}. \bibinfo{pages}{325--334}.
\newblock


\bibitem[\protect\citeauthoryear{Zhang, Lian, and Yang}{Zhang
  et~al\mbox{.}}{2017}]%
        {zhang2017discrete}
\bibfield{author}{\bibinfo{person}{Yan Zhang}, \bibinfo{person}{Defu Lian},
  {and} \bibinfo{person}{Guowu Yang}.} \bibinfo{year}{2017}\natexlab{}.
\newblock \showarticletitle{Discrete personalized ranking for fast
  collaborative filtering from implicit feedback}. In
  \bibinfo{booktitle}{\emph{AAAI}}, Vol.~\bibinfo{volume}{31}.
\newblock


\bibitem[\protect\citeauthoryear{Zhang and Zhu}{Zhang and Zhu}{2019}]%
        {zhang2019doc2hash}
\bibfield{author}{\bibinfo{person}{Yifei Zhang} {and} \bibinfo{person}{Hao
  Zhu}.} \bibinfo{year}{2019}\natexlab{}.
\newblock \showarticletitle{Doc2hash: Learning discrete latent variables for
  documents retrieval}. In \bibinfo{booktitle}{\emph{ACL}}.
  \bibinfo{pages}{2235--2240}.
\newblock


\bibitem[\protect\citeauthoryear{Zhang and Zhu}{Zhang and Zhu}{2020}]%
        {zhang2020discrete}
\bibfield{author}{\bibinfo{person}{Yifei Zhang} {and} \bibinfo{person}{Hao
  Zhu}.} \bibinfo{year}{2020}\natexlab{}.
\newblock \showarticletitle{Discrete Wasserstein Autoencoders for Document
  Retrieval}. In \bibinfo{booktitle}{\emph{ICASSP}}. IEEE,
  \bibinfo{pages}{8159--8163}.
\newblock


\bibitem[\protect\citeauthoryear{Zhu, Long, Wang, and Cao}{Zhu
  et~al\mbox{.}}{2016}]%
        {zhu2016deep}
\bibfield{author}{\bibinfo{person}{Han Zhu}, \bibinfo{person}{Mingsheng Long},
  \bibinfo{person}{Jianmin Wang}, {and} \bibinfo{person}{Yue Cao}.}
  \bibinfo{year}{2016}\natexlab{}.
\newblock \showarticletitle{Deep hashing network for efficient similarity
  retrieval}. In \bibinfo{booktitle}{\emph{AAAI}}, Vol.~\bibinfo{volume}{30}.
\newblock


\end{thebibliography}
}
\newpage
\clearpage
\appendix
\setcounter{table}{0}
\setcounter{figure}{0}

\begin{table}[th]
\setlength{\abovecaptionskip}{0.2cm}
\setlength{\belowcaptionskip}{0.2cm}
\caption {Notations and meanings. }
\vspace{-0.05in}
\label{tab:notation}
  \footnotesize
  \begin{tabular}{c|l} 
     \hline
          {\bf Notation} & {\bf Explanation}\\
    \hline\hline
      $d$, $L$    & {Embedding dimensions and graph convolution layers.} \\
    \hline
      $\mathcal{U}$, $\mathcal{I}$ & Collection of users and items. \\
    \hline
       $\mathcal{N}(x)$ & {Neighbors of node $x$.} \\
    \hline
          $\boldsymbol{v}_{x}^{(l)}$ & {Full-precision embedding of node $x$ at $l$-th convolution.}\\
    \hline
          $\boldsymbol{q}_{x}^{(l)}$ & {Binarized embedding of of node $x$ at $l$-th quantization.}\\
    \hline
        $\alpha_{x}^{(l)}$ & {$l$-th embedding scaler of node $x$.}\\
    \hline
        $\mathcal{A}_x$ and $\mathcal{Q}_x$ & {Binarized embedding table of $x$ learned by BiGeaR. } \\
    \hline
        $w_l$   & {$l$-th weight in predicting matching score.} \\
    \hline
         $y_{u,i}$     & {A scalar indicates the existence of user-item interaction.} \\
    \hline 
      $\widehat{y}^{tch}_{u,i}$  &   {Predicted score based on full-precision embeddings.}\\
    \hline
    $\widehat{y}^{std}_{u,i}$  &   {Predicted score based on binarized embeddings.}\\
    \hline
      $\emb{\widehat{y}}^{tch,\,(l)}_{u}$  & {Predicted scores of $u$ based on $l$-th embeddings segments.}\\
    \hline
      $\emb{\widehat{y}}^{std,\,(l)}_{u}$  & {Predicted scores of $u$ based on $l$-th quantized segments.}\\
    \hline 
      $S_{tch}^{(l)}(u)$ & {pseudo-positive training samples of $u$.}\\
    \hline
      $w_k$   &   {$k$-th weight in inference distillation loss.}\\
    \hline     
      $\mathcal{L}_{BPR}^{tch}$, $\mathcal{L}_{BPR}^{std}$  & {BPR loss based on full-precision and binarized scores.} \\
    \hline     
      $\mathcal{L}_{ID}$  & {Inference distillation loss.} \\
    \hline     
      $\mathcal{L}$ & {Objective function of BiGeaR.} \\
    \hline
      $u(\cdot)$, $\delta(\cdot)$ & Unit-step function and Dirac delta function.\\
    \hline
      $\lambda$, $\lambda_1$, $\lambda_2$, $\gamma$, $\eta$ & {Hyper-parameters and the learning rate.} \\
    \hline
  \end{tabular}
\end{table}

\section{Notation Table}
\label{app:notation}
We list key notations in Table~\ref{tab:notation}.

\begin{figure*}[tbh]
\begin{minipage}{1\textwidth}
\hspace{-0.05in}
\includegraphics[width=7in]{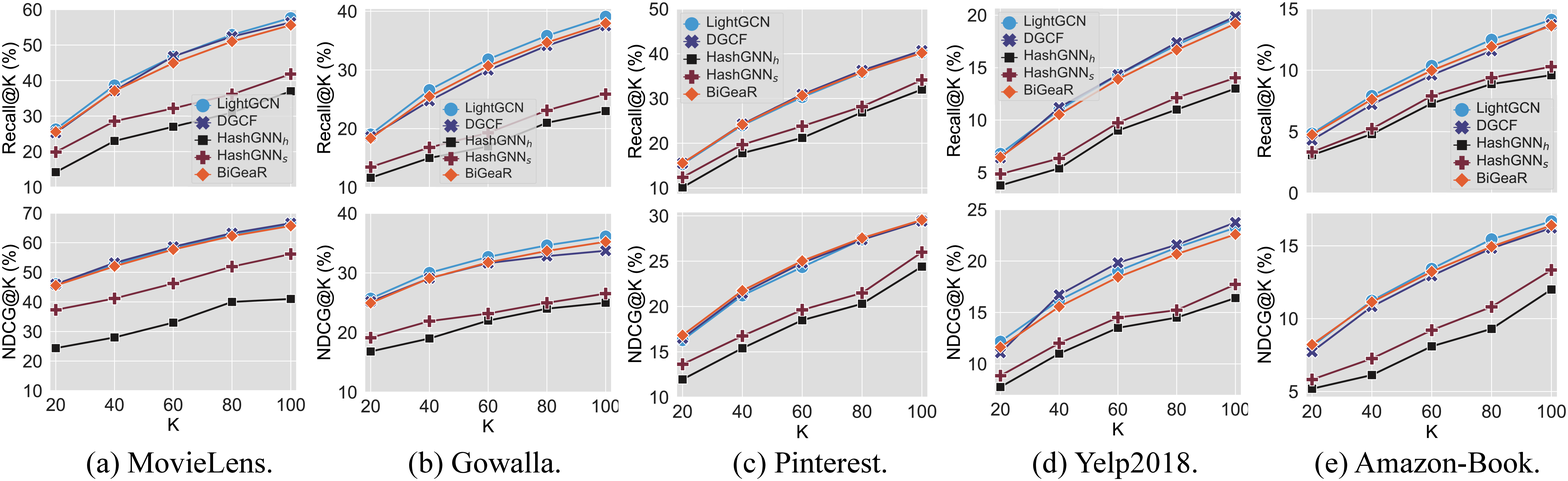}
\end{minipage} 
\vspace{-0.05in}
\caption{Top-K recommendation curve.}
\label{fig:topk}
\end{figure*}

\section{Pseudo-codes of BiGeaR}
\label{app:algo}
The pseudo-codes of BiGeaR are attached in Algorithm~\ref{alg:bigear}.

\begin{algorithm}[t]
\footnotesize
\caption{BiGeaR algorithm.}
\label{alg:bigear}
\LinesNumbered  
\KwIn{Interaction graph; trainable embeddings {\footnotesize $\boldsymbol{v}_{\{\cdots\}}$}; hyper-parameters: {\footnotesize $L$, $\eta$, $\lambda$, $\lambda_1$, $\lambda_2$, $\gamma$. } }
\KwOut{Prediction function $\mathcal{F}(u,i)$} 
$\mathcal{A}_u \gets \emptyset$, $\mathcal{A}_i \gets \emptyset$, $\mathcal{Q}_u \gets \emptyset$, $\mathcal{Q}_i \gets \emptyset$;\\
\While{\rm{BiGeaR not converge}}{
    \For{$l = 1, \cdots, L$}{
         $\boldsymbol{v}^{(l)}_u \gets\sum_{i\in \mathcal{N}(u)} \frac{1}{\sqrt{|\mathcal{N}(u)|\cdot|\mathcal{N}(i)|}}\boldsymbol{v}^{(l-1)}_i$, \\ 
         $\boldsymbol{v}^{(l)}_i\gets \sum_{u\in \mathcal{N}(i)} \frac{1}{\sqrt{|\mathcal{N}(i)|\cdot|\mathcal{N}(u)|}}\boldsymbol{v}^{(l-1)}_u$. \\
          \If{{\color{blue} with inference distillation}}{
              $\boldsymbol{q}_u^{(l)}\gets \sign\big(\boldsymbol{v}^{(l)}_u\big), \ \ \boldsymbol{q}_i^{(l)}\gets \sign\big(\boldsymbol{v}^{(l)}_i\big)$, \\
              $\alpha_u^{(l)} \gets \frac{||\boldsymbol{V}_u^{(l)}||_1}{d}$, $\alpha_i^{(l)} \gets \frac{||\boldsymbol{V}_i^{(l)}||_1}{d}$; \\
              Update ($\mathcal{A}_u$, $\mathcal{Q}_u$), ($\mathcal{A}_i$, $\mathcal{Q}_i$) with $\alpha_u^{(l)}\boldsymbol{q}_u^{(l)}$, $\alpha_i^{(l)}\boldsymbol{q}_i^{(l)}$; \\              
          }
        }

     $\widehat{y}^{\,tch}_{u,i}\gets \Big<\big|\big|_{l=0}^L w_l\boldsymbol{v}_u^{(l)}, \big|\big|_{l=0}^L w_l\boldsymbol{v}_i^{(l)}\Big>$. \\

      \If{{\color{blue} with inference distillation}}{
            $\boldsymbol{q}_u^{(0)} \gets \sign\big(\boldsymbol{v}^{(0)}_u\big), \ \ \boldsymbol{q}_i^{(0)} \gets \sign\big(\boldsymbol{v}^{(0)}_i\big)$, \\
              $\alpha_u^{(0)} \gets \frac{||\boldsymbol{V}_u^{(0)}||_1}{d}$, $\alpha_i^{(0)} \gets \frac{||\boldsymbol{V}_i^{(0)}||_1}{d}$; \\
              Update ($\mathcal{A}_u$, $\mathcal{Q}_u$), ($\mathcal{A}_i$, $\mathcal{Q}_i$) with $\alpha_u^{(0)}\boldsymbol{q}_u^{(0)}$, $\alpha_i^{(0)}\boldsymbol{q}_i^{(0)}$;
            $\widehat{y}^{\,std}_{u,i} =  \big<f(\mathcal{A}_u, \mathcal{Q}_u), f(\mathcal{A}_i, \mathcal{Q}_i)\big>$; \\ 
             $\{\widehat{y}^{tch,\,(l)}_{u,i}\}_{l=0, 1, \cdots, L} \gets \text{get score segments from} \widehat{y}^{\,tch}_{u,i}$; \\ 
             $\{\widehat{y}^{std,\,(l)}_{u,i}\}_{l=0, 1, \cdots, L} \gets \text{get score segments from}  \widehat{y}^{\,std}_{u,i}$; \\ 
              $\mathcal{L}_{ID} \gets$ compute loss with {\scriptsize $\{\widehat{y}^{tch,\,(l)}_{u,i}\}_{l=0, 1, \cdots, L}$, $\{\widehat{y}^{std,\,(l)}_{u,i}\}_{l=0, 1, \cdots, L}$}. \\	
             $\mathcal{L} \gets$ compute $\mathcal{L}^{std}_{BPR} and \mathcal{L}_{ID}$. \\	

      }
      \Else{
            $\mathcal{L} \gets$ compute $\mathcal{L}^{tch}_{BPR}$. \\ 	
      }

       Optimize BiGeaR with regularization;\\ 

}
\KwRet $\mathcal{F}$.\\
\end{algorithm}

\section{Datasets}
\label{app:dataset}
\begin{itemize}[leftmargin=*]
\item \textbf{MovieLens}~\cite{hashgnn,he2016fast,chen2021modeling,chen2021attentive} is a widely adopted benchmark for movie recommendation. Similar to the setting in~\cite{hashgnn,he2016fast,chen2021modeling}, $y_{u,i} = 1$ if user $u$ has an explicit rating score towards item $i$, otherwise $y_{u,i} = 0$. In this paper, we use the MovieLens-1M data split.

\item \textbf{Gowalla}~\cite{ngcf,hashgnn,lightgcn,dgcf} is the check-in dataset~\cite{liang2016modeling} collected from Gowalla, where users share their locations by check-in. To guarantee the quality of the dataset, we extract users and items with no less than 10 interactions similar to~\cite{ngcf,hashgnn,lightgcn,dgcf}. 

\item \textbf{Pinterest}~\cite{geng2015learning,hashgnn} is an implicit feedback dataset for image recommendation~\cite{geng2015learning}. Users and images are modeled in a graph. Edges represent the pins over images initiated by users. In this dataset, each user has at least 20 edges. 

\item \textbf{Yelp2018}~\cite{ngcf,dgcf,lightgcn} is collected from Yelp Challenge 2018 Edition, where local businesses such as restaurants are treated as items. We retain users and items with over 10 interactions similar to~\cite{ngcf,dgcf,lightgcn}.

\item \textbf{Amazon-Book}~\cite{ngcf,dgcf,lightgcn} is organized from the book collection of Amazon-review for product recommendation~\cite{he2016ups}. Similarly to~\cite{ngcf,lightgcn,dgcf}, we use the 10-core setting to graph nodes.  
\end{itemize} 

\section{Competing Methods}
\label{app:method}
\begin{itemize}[leftmargin=*,topsep=0.5pt,parsep=0.5pt]
\item \textbf{LSH}~\cite{lsh} is a representative hashing method to approximate the similarity search for massive high-dimensional data. We follow the adaptation in~\cite{hashgnn} to it for Top-K recommendation. 

\item \textbf{HashNet}~\cite{hashnet} is a state-of-the-art deep hashing method that is originally proposed for multimedia retrieval tasks.
We use the same adaptation strategy in~\cite{hashgnn} to it for recommendation.

\item \textbf{CIGAR}~\cite{kang2019candidate} is a hashing-based method for fast item candidate generation, followed by complex full-precision re-ranking algorithms. We use its quantization part for fair comparison. 

\item \textbf{GumbelRec} is a variant of our model with the implementation of Gumbel-softmax for categorical variable quantization~\cite{gumbel1,gumbel2,zhang2019doc2hash}.
GumbelRec utilizes the Gumbel-softmax trick to replace $\sign(\cdot)$ function for embedding binarization.  

\item \textbf{HashGNN}~\cite{hashgnn} is the state-of-the-art end-to-end 1-bit quantization recommender system. 
\textbf{HashGNN$_h$} denotes its vanilla \textit{hard encoding} version; and \textbf{HashGNN$_s$} is the relaxed version of replacing several quantized digits with the full-precision ones.

\item \textbf{NeurCF}~\cite{neurcf} is a classical neural network model to capture user-item nonlinear feature interactions for collaborative filtering.

\item \textbf{NGCF}~\cite{ngcf} is a state-of-the-art graph-based collaborative filtering model that largely follows the standard GCN~\cite{kipf2016semi}. 

\item \textbf{DGCF}~\cite{dgcf} is one of the latest graph-based model that learns disentangled user intents for better Top-K recommendation.

\item \textbf{LightGCN}~\cite{lightgcn} is another latest GCN-based recommender system that presents a more concise and powerful model structure 
with state-of-the-art performance.
\end{itemize}

\section{Hyper-parameter Settings}
\label{app:parameter}
We report all hyper-parameter settings in Table~\ref{tab:hyperparameter}.

\begin{table}[thb]
\setlength{\abovecaptionskip}{0.2cm}
\setlength{\belowcaptionskip}{0.2cm}
\centering
\small
\caption{Hyper-parameter settings for the five datasets.}
\vspace{-0.05in}
\label{tab:hyperparameter}
\setlength{\tabcolsep}{1.5mm}{
\begin{tabular}{c | c  c  c  c  c}
\toprule
                  & MovieLens   & Gowalla    & Pinterest     & Yelp2018  & Amazon-Book\\
\midrule 
\midrule
  $B$               &  2048       &   2048     & 2048        &  2048   &  2048  \\
  $d$         &  256        &   256      &  256    & 256  & 256   \\
  $\eta$            & $1\times10^{-3}$    & $1\times10^{-3}$  & $5\times10^{-4}$      & $5\times10^{-4}$        & $5\times10^{-4}$      \\
  $\lambda$         & $1\times10^{-4}$    & $5\times10^{-5}$  & $1\times10^{-4}$      & $1\times10^{-4}$        & $1\times10^{-6}$      \\
  $\lambda_1$   & 1 & 1 & 1 & 1 & 1\\
  $\lambda_2$   &  0.1 & 0.1 & 0.1 & 0.1 & 0.1 \\
  $\gamma$      & 1  & 1 & 1 & 1 & 1\\
  $L$       & 2 & 2 & 2 & 2 & 2 \\
\bottomrule
\end{tabular}}
\end{table}

\section{Additional Experimental Results}

\subsection{Top-K Recommendation Curve}
\label{app:topk} 
We curve the Top-K recommendation by varying K from 20 to 100 and compare BiGeaR with several selected models.
As shown in Figure~\ref{fig:topk}, BiGeaR consistently presents the performance superiority over HashGNN, and shows the competitive recommendation accuracy with DGCF and LightGCN.

\subsection{Implementation of Embedding Scaler ${\alpha}^{(l)}$}
\label{app:scaler}

We set the embedding scaler to learnable (denoted by \textsl{LB}) and show the results in Table~\ref{tab:scaler}.
We observe that, the design of learnable embedding scaler does not achieve the expected performance. 
This is probably because there is no direct mathematical constraint to it and thus the parameter search space is too large to find the optimum by stochastic optimization.

\begin{table}[h]
\setlength{\abovecaptionskip}{0.2cm}
\setlength{\belowcaptionskip}{0.2cm}
\centering
\footnotesize
\caption{Implementation of Embedding Scaler.}
\vspace{-0.05in}
\label{tab:scaler}
\setlength{\tabcolsep}{0.65mm}{
\begin{tabular}{c |c c|c c|c c|c c|c c}
\toprule
 ~ & \multicolumn{2}{c|}{MovieLens} & \multicolumn{2}{c|}{Gowalla} & \multicolumn{2}{c|}{Pinterest} & \multicolumn{2}{c|}{Yelp2018} & \multicolumn{2}{c}{Amazon-book} \\
               ~  & R@20 & N@20 & R@20 & N@20 & R@20 & N@20 & R@20 & N@20 & R@20 & N@20\\
\midrule
\midrule[0.1pt]
  \multirow{2}*{\footnotesize \textsl{LB}}    &{23.07} & {41.42}   & {17.01}& {23.11}  & {14.19}& {15.29}  & {6.05} & {10.80} & {4.52} & {7.85}\\
  ~        &\textit{\color{blue} \scriptsize{-9.78\%}}  &\textit{\color{blue} \scriptsize{-9.09\%}}  &\textit{\color{blue} \scriptsize{-7.35\%}}  &\textit{\color{blue} \scriptsize{-7.41\%}}  &\textit{\color{blue} \scriptsize{-8.86\%}}  &\textit{\color{blue} \scriptsize{-9.15\%}}  &\textit{\color{blue} \scriptsize{-6.49\%}}  &\textit{\color{blue} \scriptsize{-6.90\%}} &\textit{\color{blue} \scriptsize{-3.42\%}}  &\textit{\color{blue}  \scriptsize{-3.33\%}}\\
  \midrule[0.1pt]
\cellcolor{best}{\footnotesize \textbf{BiGeaR} }  &\textbf{25.57}& \textbf{45.56}   & \textbf{18.36}& \textbf{24.96}  & \textbf{15.57}& \textbf{16.83}  & \textbf{6.47}& \textbf{11.60} & \textbf{4.68}& \textbf{8.12}\\
\bottomrule
\end{tabular}}
\end{table}

\subsection{\textbf{Implementation of $w_l$.}}
\label{app:wl}
We try the following three additional implementation of $w_l$ and report the results in Tables~\ref{tab:wl}.
\begin{enumerate}[leftmargin=*]
\item {\small $w_l$ = $\frac{1}{L+1}$} equally contributes for all embedding segments.

\item {\small $w_l$ = $\frac{1}{L+1-l}$} is positively correlated to the $l$ value, so as to highlight higher-order structures of the interaction graph. 

\item {\small $w_l$ = $2^{-(L+1-l)}$} is positively correlated to $l$ with exponentiation.
\end{enumerate}
The experimental results show that implementation (2) performs fairly well compared to the others, demonstrating the importance of highlighting higher-order graph information.
This corroborates the design of our implementation in BiGeaR, i.e., {\small $w_l$ $\propto$ $l$}, which however is simpler and effective with better recommendation accuracy.

\begin{table}[h]
\setlength{\abovecaptionskip}{0.2cm}
\setlength{\belowcaptionskip}{0.2cm}
\centering
\footnotesize
\caption{Implementation of $w_l$.}
\vspace{-0.05in}
\label{tab:wl}
\setlength{\tabcolsep}{0.85mm}{
\begin{tabular}{c | c c | c c | c c | c c | c c }
\toprule
~  & \multicolumn{2}{c|}{MovieLens} & \multicolumn{2}{c|}{Gowalla} & \multicolumn{2}{c|}{Pinterest} & \multicolumn{2}{c|}{Yelp2018} & \multicolumn{2}{c}{Amazon-Book}\\
               ~  & R@20 & N@20 & R@20 & N@20 & R@20 & N@20 & R@20 & N@20 &R@20 & N@20\\
\midrule[0.1pt]
\midrule[0.1pt]
  (1)           &{22.75}  &{41.13} &{16.15} &{21.82} &{14.16} &{15.48} &{5.88} &{10.32} &{4.46} &{7.63} \\
  (2)           &{25.07}  &{44.64} &{17.81} &{24.46} &{15.26} &{16.57} &{6.40} &{11.38} &{4.58} &{7.96} \\        
  (3)     &{21.23}  &{37.81} &{15.24} &{20.71} &{12.93} &{14.28} &{5.24} &{9.51} &{3.74} &{64.98} \\  
\midrule[0.1pt]
\cellcolor{best}\textbf{Best}     &{\textbf{25.57}}  &{\textbf{45.56}} &{\textbf{18.36}} &{\textbf{24.96}} &{\textbf{15.57}} &{\textbf{16.83}} &{\textbf{6.47}} &{\textbf{11.60}}  &{\textbf{4.68}}  &{\textbf{8.12}}  \\ 
      
\bottomrule
\end{tabular}}
\end{table}

\subsection{\textbf{Implementation of $w_k$.}}
\label{app:wk}
We further evaluate different $w_k$:
\begin{enumerate}[leftmargin=*]
\item  {\small $w_k$ = $\frac{R-k}{R}$} is negatively correlated to the ranking position $k$.
\item  {\small $w_k$ = $\frac{1}{k}$} is inversely proportional to position $k$.
\item  {\small $w_k$ = $2^{-k}$} is exponential to the value of $-k$.
\end{enumerate}
We observe from Table~\ref{tab:wk} that the implementation (3) works slightly worse than Equation~(\ref{eq:wk}) but generally better than the other two methods. 
This show that the exponential modeling is more effective to depict the importance contribution of items for approximating the tailed item popularity~\cite{rendle2014improving}.
Moreover, Equation~(\ref{eq:wk}) introduces hyper-parameters to provide the flexibility of adjusting the function properties for different datasets.

\begin{table}[h]
\setlength{\abovecaptionskip}{0.2cm}
\setlength{\belowcaptionskip}{0.2cm}
\centering
\footnotesize
\caption{Implementation of $w_k$.}
\vspace{-0.05in}
\label{tab:wk}
\setlength{\tabcolsep}{0.86mm}{
\begin{tabular}{c | c c | c c | c c | c c | c c }
\toprule
~ & \multicolumn{2}{c|}{MovieLens} & \multicolumn{2}{c|}{Gowalla} & \multicolumn{2}{c|}{Pinterest} & \multicolumn{2}{c|}{Yelp2018} & \multicolumn{2}{c}{Amazon-Book}\\
               ~  & R@20 & N@20 & R@20 & N@20 & R@20 & N@20 & R@20 & N@20 &R@20 & N@20\\
\midrule[0.1pt]
\midrule[0.1pt]
  (1)           &{24.97}  &{44.33} &{17.96} &{24.87} &{15.11} &{16.20} &{6.28} &{11.21} &{4.43} &{7.78} \\
  (2)           &{25.08}  &{45.19} &{17.95} &{24.95} &{15.18} &{16.34} &{6.27} &{11.25} &{4.48} &{7.92} \\        
  (3)     &{25.16}  &{44.92} &{18.32} &{24.81} &{15.26} &{16.65} &{6.33} &{11.36} &{4.53} &{8.06} \\  
\midrule[0.1pt]
\cellcolor{best}\textbf{Best}     &{\textbf{25.57}}  &{\textbf{45.56}} &{\textbf{18.36}} &{\textbf{24.96}} &{\textbf{15.57}} &{\textbf{16.83}} &{\textbf{6.47}} &{\textbf{11.60}}  &{\textbf{4.68}}  &{\textbf{8.12}}  \\ 
      
\bottomrule
\end{tabular}}
\end{table}



\end{document}